\documentclass[trackchanges]{aastex631}

\submitjournal{AJ}
\shorttitle{The DESI DR1 Fork Geometry}
\shortauthors{Nigoche-Netro et al.}

\usepackage{amsmath}
\usepackage{graphicx}

\begin{document}

\title{The ``Fork'' Geometry in the DESI Main Survey DR1: Implications of Target Selection for Galaxy Evolution Studies}

\correspondingauthor{A. Nigoche-Netro}
\email{alberto.nigoche@academicos.udg.mx}

\author[0000-0002-8683-0982]{A. Nigoche-Netro}
\affiliation{Instituto de Astronom\'ia y Meteorolog\'ia,
CUCEI, Universidad de Guadalajara, 
Guadalajara, Jal. 44130, M\'exico.}

\author[0000-0002-2321-8657]{P. Lagos}
\affiliation{Instituto de Astrof\'isica e Ci\^encias do
Espa\c{c}o, Universidade do Porto, CAUP, \\
Rua das Estrelas, 4150-762 Porto, Portugal.}
\affiliation{Institute of Astrophysics, Facultad de Ciencias Exactas, 
Universidad Andr\'es Bello, \\
Sede Concepci\'on, Talcahuano, Chile.}

\author[0000-0001-9716-5335]{R. J. Diaz}
\affiliation{Gemini Observatory, NSF NOIRLab
950 N Cherry Ave, Tucson AZ, USA.}
\affiliation{Universidad Nacional de C\'ordoba
Laprida 854, C\'ordoba, CPA: X5000BGR,\\ Argentina.}

\author[0000-0002-3449-1237]{E. Garc\'ia-Manzan\'arez}
\affiliation{Instituto de Astronom\'ia y Meteorolog\'ia,
CUCEI, Universidad de Guadalajara, 
Guadalajara, Jal. 44130, M\'exico.}

\author{J. A. Quezada-Pérez}
\affiliation{Instituto de Astronom\'ia y Meteorolog\'ia,
CUCEI, Universidad de Guadalajara, 
Guadalajara, Jal. 44130, M\'exico.}

\author{R. Ibarra-Nuño}
\affiliation{Instituto de Astronom\'ia y Meteorolog\'ia,
CUCEI, Universidad de Guadalajara, 
Guadalajara, Jal. 44130, M\'exico.}


\begin{abstract}

The Dark Energy Spectroscopic Instrument (DESI) Main Survey Data Release 1 (DR1) provides an unprecedented spectroscopic view of galaxies and quasars across a large fraction of the observable Universe. However, the survey's tracer-dependent targeting strategy introduces complex observational selection effects that strongly influence the apparent distribution of galaxies in the redshift--luminosity plane. In this work, we investigate the origin of the characteristic ``fork'' geometry observed in DESI DR1 and show that the observed distribution is consistent with an interpretation in which it arises from the combined action of intrinsic galaxy bimodality and the survey's redshift-dependent selection function. We show that the observed underdensity separating the dominant galaxy populations is influenced not only by the physical Green Valley but also by survey selection effects, since it is further shaped by magnitude limits, surface-brightness selection, and structural incompleteness. Using DESI DR1 spectroscopy together with Legacy Survey photometry, we analyze the roles of the BGS, LRG, ELG, and QSO target classes in producing the observed distribution and assess their impact on luminosity-function measurements. Our results indicate that both target selection and observational biases significantly affect the observed galaxy distribution and the evolutionary trends inferred from it. We summarize practical observational considerations aimed at minimizing these effects and enabling more robust studies of galaxy evolution with DESI.

\end{abstract}


\section{Introduction} \label{sec:intro}

The Dark Energy Spectroscopic Instrument (DESI) is currently conducting the most ambitious spectroscopic survey to date, aimed at providing sub-percent precision measurements of the expansion history of the Universe through Baryon Acoustic Oscillations (BAO) and Redshift-Space Distortions (RSD) \citep{DESI2016, DESI2024}. To achieve these cosmological goals, DESI targets multiple extragalactic tracers optimized for different redshift ranges. The DR1 marks a milestone, offering a dense spectroscopic view of the Universe supplemented by deep multi-band imaging from the Legacy Surveys \citep{Dey2019}.

However, a fundamental challenge arises for researchers utilizing these data for galaxy formation and evolution studies: the survey's target selection logic is primarily designed for cosmological efficiency, not population continuity. When examining the distribution of galaxies in the redshift--absolute magnitude ($z$-$M$) plane across the DR1 sample, a prominent fork-like structure emerges. This structure is characterized by a stark bifurcation at $z < 0.7$ and an apparent convergence of populations at higher redshifts ($z \gtrsim 0.7$).

These features do not solely reflect the intrinsic galaxy population, but are strongly shaped by the discrete DESI targeting strategies. Selection effects in flux-limited galaxy surveys have been widely discussed in the literature \citep[e.g.,][]{Nigoche2022, DESI2024}. Without a rigorous characterization of these selection gaps and the biases associated with the ``scaffolded'' redshift design of the DR1, evolutionary analyses risk identifying spurious trends in quenching timescales and mass assembly.

Throughout this paper, all photometric quantities are expressed in the AB magnitude system \citep{Oke1983}. Where synthetic stellar populations or stellar mass estimates are referenced (e.g., in spectral fitting performed by \texttt{FastSpecFit}), a \citet{Chabrier2003} Initial Mass Function (IMF) is assumed. We adopt a flat $\Lambda$CDM cosmological model with $H_0 = 67.4~\mathrm{km~s^{-1}~Mpc^{-1}}$, $\Omega_m = 0.31$, and $\Omega_\Lambda = 0.69$, following the results of \citet{Planck2020}. Unless explicitly noted otherwise, $M$ denotes rest-frame absolute magnitude.

This paper dissects the interplay between intrinsic galaxy bimodality \citep[e.g.,][]{Strateva2001, Kauffmann2003, Baldry2004, Faber2007, Schawinski2014} and instrumental target selection in the DESI DR1 dataset \citep{Myers2023, Hahn2023, Zhou2023, Raichoor2023, Chaussidon2023}. The structure of the paper is as follows: In Section~\ref{sec:fork} we describe the anatomy of the fork geometry and its physical and instrumental origins. Section~\ref{sec:data} presents the DESI DR1 and Legacy Surveys data products and baseline corrections. Section~\ref{sec:mitigation} summarizes practical considerations for mitigating observational biases. Section~\ref{sec:lf_analysis} provides a quantitative diagnostic of these distortions using luminosity functions. Section~\ref{sec:discussion} discusses the implications for galaxy evolution studies, and Section~\ref{sec:conclusions} summarizes our main conclusions.

\section{The Anatomy of the Fork Geometry} \label{sec:fork}

In this section, we describe the observed fork geometry in the DESI DR1 data and establish its dual origin: physical galaxy bimodality and instrumental selection.

\subsection{Physical Bimodality and its Redshift Dependence}

Galaxy populations are naturally bimodal, divided into the star-forming ``Blue Cloud'' and the quiescent ``Red Sequence'' \citep{Strateva2001, Bell2004}. This bimodality is a fundamental function of cosmic time; as the Universe evolves, galaxies migrate from the Blue Cloud to the Red Sequence through various quenching mechanisms, including AGN feedback, environmental stripping, and gas exhaustion \citep{Faber2007, Schawinski2014}. The Green Valley is commonly interpreted as a transitional region between the Blue Cloud and the Red Sequence.

In an unbiased $z$--$M$ diagram, we expect to see a continuous but bimodal distribution where the Red Sequence gradually populates toward $z = 0$ \citep{Weiner2005}. However, as we demonstrate below, the DESI selection function significantly modifies the visibility of this transition, enhancing the apparent depletion of galaxies in the region commonly associated with the Green Valley.

\subsection{DESI Target Selection Effects}

The apparent bimodal ``fork'' structure observed in the DESI DR1 ($z$-$M$) distribution is strongly amplified by the survey targeting strategy \citep{Myers2023}, although it ultimately originates from the intrinsic bimodality of the galaxy population. Rather than tracing a uniformly selected galaxy population, DESI combines multiple spectroscopic target classes, each optimized for different redshift regimes and astrophysical tracers \citep{Ruiz2020, Zhou2023}. 

\begin{figure}[ht]
    \centering
    \includegraphics[width=0.4\textwidth]{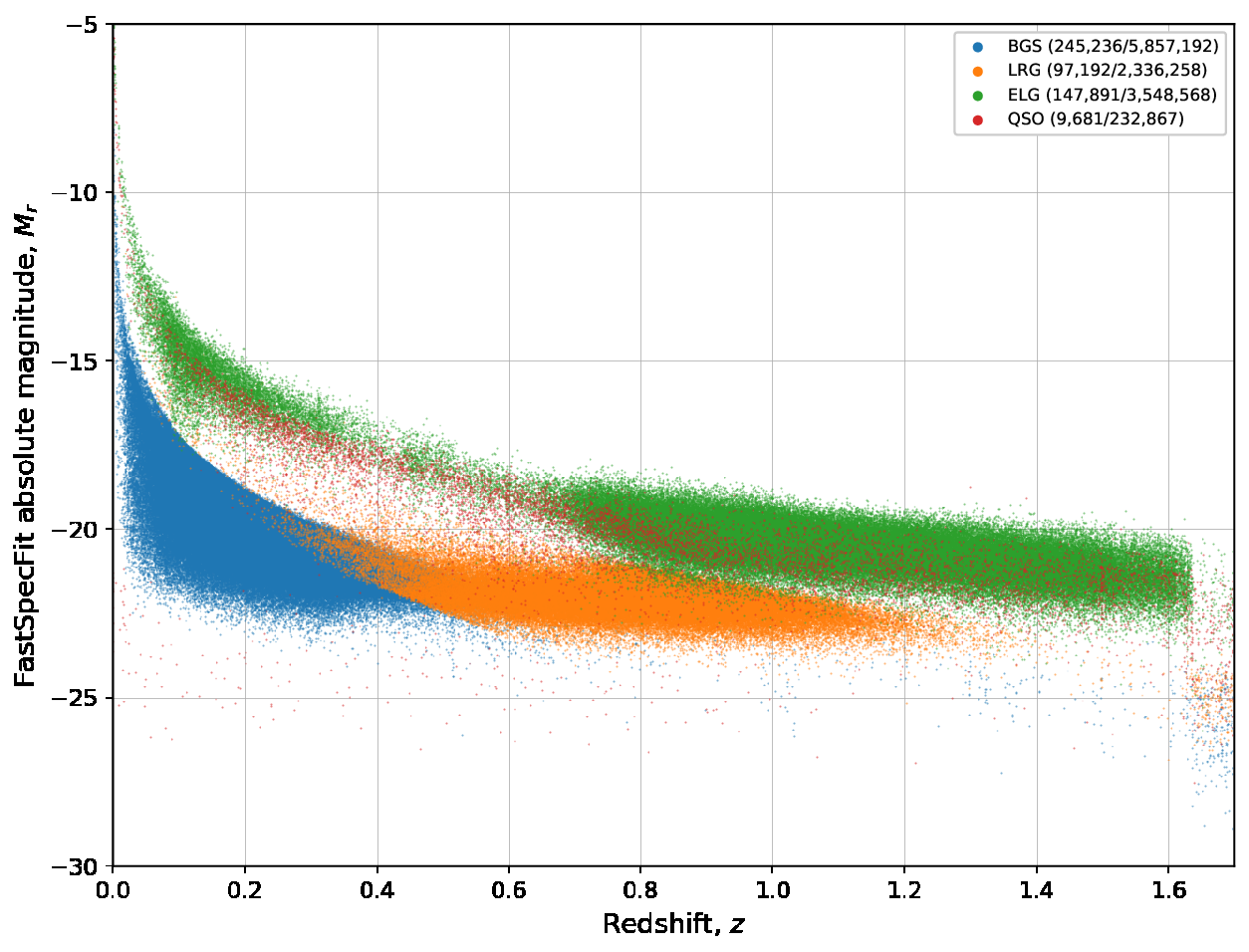}
    \caption{Redshift distribution as a function of the FastSpecFit rest-frame absolute magnitude in the SDSS $r$ band, $M_r$, for DESI DR1 galaxies. Points are grouped according to the DESI target classes, illustrating the relative location of the different spectroscopic selection populations within the $z$--$M_r$ plane. The absolute magnitudes are derived from the FastSpecFit value-added catalog and include Galactic extinction corrections, spectrophotometric $K$-corrections, and aperture normalization to the total broad-band photometric flux. For visual clarity, only a random subset of the full sample is displayed. The plotted fraction of each target class is indicated in the upper-right corner.}
    \label{fig:TargetClasseSMr}
\end{figure}

The principal target populations (see Figure~\ref{fig:TargetClasseSMr}) are:

\begin{itemize}
\item \textbf{Bright Galaxy Survey (BGS)\citep{Hahn2023}:} The BGS sample dominates the low-redshift Universe ($z \lesssim 0.4$) and primarily traces nearby star-forming and intermediate-mass galaxies. This population forms the lower branch of the observed fork structure, extending toward low luminosities at very small redshifts. The rapid depletion of faint galaxies is driven primarily by Malmquist incompleteness and surface-brightness selection effects.

\item \textbf{Luminous Red Galaxies (LRGs) \citep{Zhou2023}:} The LRG program is optimized to target massive, intrinsically luminous, quiescent galaxies over $0.4 \lesssim z \lesssim 1.1$. These objects define the continuity of the lower, high-luminosity branch of the fork. Because the LRG selection employs aggressive color--magnitude cuts to reject low-redshift contaminants, the local Red Sequence becomes artificially underrepresented at $z \lesssim 0.4$.

\item \textbf{Emission Line Galaxies (ELGs) \citep{Raichoor2023}:} The ELG sample preferentially selects blue, actively star-forming galaxies over the approximate range $0.6 \lesssim z \lesssim 1.6$. ELGs populate the star-forming upper branch and progressively overlap with the LRG selections at $z \gtrsim 0.7$.

\item \textbf{Quasars (QSOs) \citep{Chaussidon2023}:} The QSO program targets compact active galactic nuclei over a broad redshift range (typically $z \gtrsim 0.8$ and extending to $z \sim 4$ for Ly$\alpha$ forest studies) using color-space selection optimized for spectroscopic Ly$\alpha$ forest studies and black-hole accretion diagnostics. Although not representative of the normal galaxy population, QSOs occupy mainly the extreme high-luminosity and high-redshift regions of the diagram, further contributing to the apparent convergence of populations at higher redshift.

\end{itemize}

The combined superposition of these heterogeneous target classes produces the characteristic ``fork'' morphology shown in Figure~\ref{fig:TargetClasseSMr}. At low redshift, the diagram separates into a lower branch dominated by BGS and LRG galaxies and an upper branch dominated by ELG and QSO systems. Between these components, the Green Valley appears underrepresented, with its observed density further reduced by the compounded effects of discrete targeting boundaries, flux incompleteness, surface-brightness dimming, and color-selection biases.

\section{Data and Basic Corrections} \label{sec:data}

This section describes the DESI DR1 and Legacy Survey DR10 data products, the basic corrections already applied by the DESI pipeline, and provides a strategic recommendation for combining these datasets.

\subsection{DESI DR1 and Legacy Survey DR10}

The analyses presented in this work are based on the DESI DR1 spectroscopic sample and the associated Legacy Survey DR10 photometry. Within the DESI framework, broad-band imaging and fiber spectroscopy provide complementary observational regimes optimized for different scientific applications. 

The DR10 Legacy Surveys photometry is generally better suited for estimating integrated luminosities and stellar masses of extended galaxies, since it is not limited by the fixed 1.5$''$ DESI fiber aperture. Conversely, DR1 spectroscopy enables accurate redshift measurements and detailed studies of stellar populations, star-formation activity, and galaxy evolution, although aperture-dependent biases may remain important for nearby extended systems. 

For studies combining DESI imaging and spectroscopy,
DR10 may provide a more suitable basis for defining
photometric samples and evaluating completeness,
whereas DR1 offers the spectroscopic information
required for accurate redshift determinations and other
spectroscopy-based analyses. Within this context, the observational considerations discussed in Section~4 are intended as practical guidelines that may help mitigate residual selection effects in galaxy evolution studies.

\subsection{Basic Pipeline Corrections}

For all DESI DR1 magnitudes used in this work, we adopt the \texttt{FastSpecFit} Value-Added Catalog values \citep{Moustakas2023}. Specifically, the \texttt{FastSpecFit} pipeline:

\begin{itemize}
    \item Corrects for Galactic foreground extinction on an object-by-object basis using the infrared dust emission maps of \citet{Schlegel1998}, adopting the updated reddening calibration from \citet{Schlafly2011} and the extinction law from \citet{Fitzpatrick1999};
    
    \item Incorporates the $K$-correction through comprehensive SED fitting by shifting stellar population models to the rest-frame;
    
    \item Circumvents local aperture losses within the $1.5''$ DESI fibers by scaling the spectroscopic flux to total broad-band imaging photometry, thereby providing a robust measure of the total integrated luminosity. We note that this scalar aperture correction accounts solely for the total integrated flux losses within the fiber aperture relative to the total photometric magnitude. Consequently, it does not account for radial color gradients, spatially resolved population variations, or complex internal light distributions across the galaxy profile—structural effects that require detailed radial surface-brightness modeling as revisited in Section~\ref{sec:SBcorrection}.


\end{itemize}

These corrections form the baseline upon which we build our unified mitigation framework in the following section.

\begin{figure}[ht]
    \centering

    \gridline{
        \fig{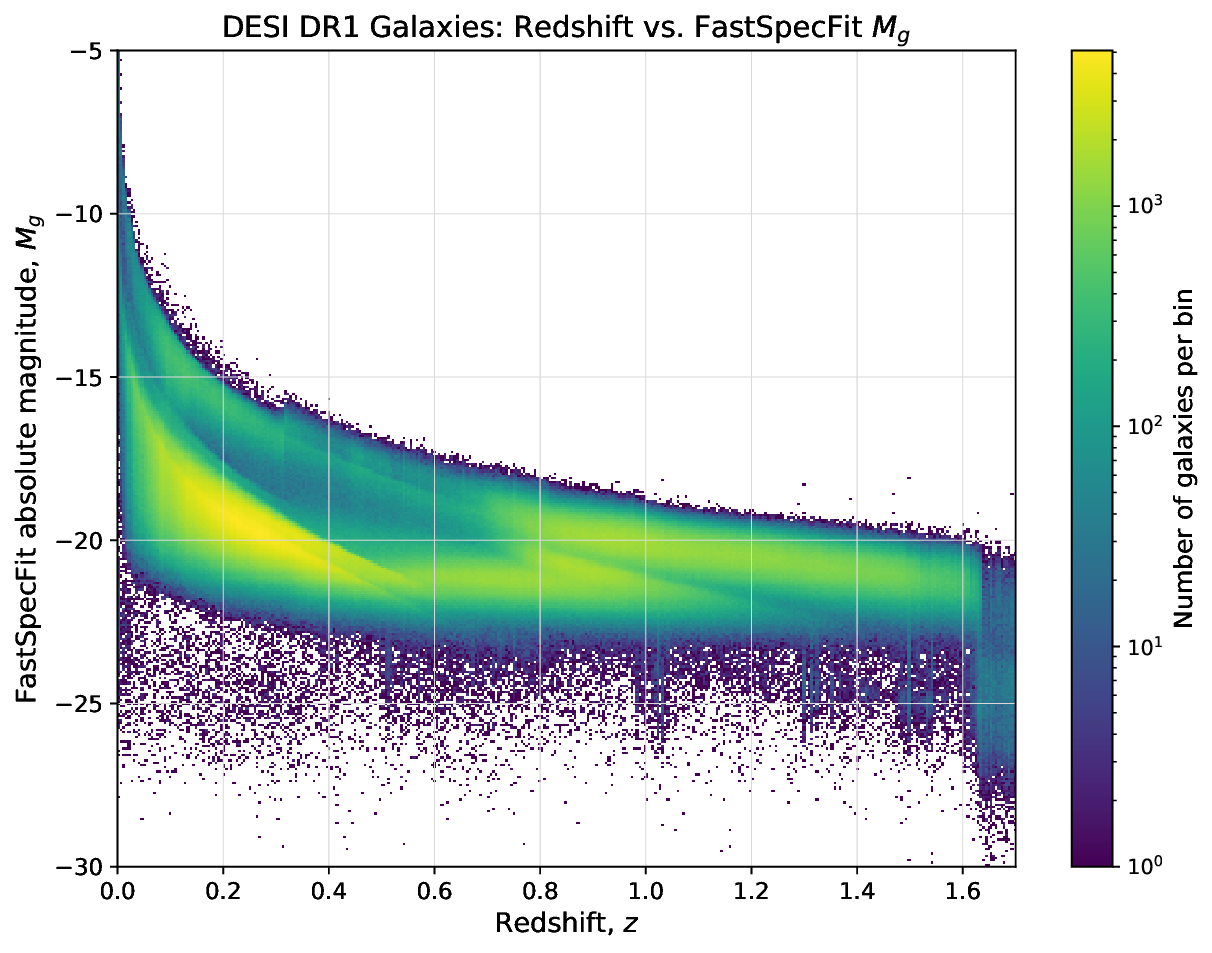}
            {0.4\textwidth}{(a)}
        \fig{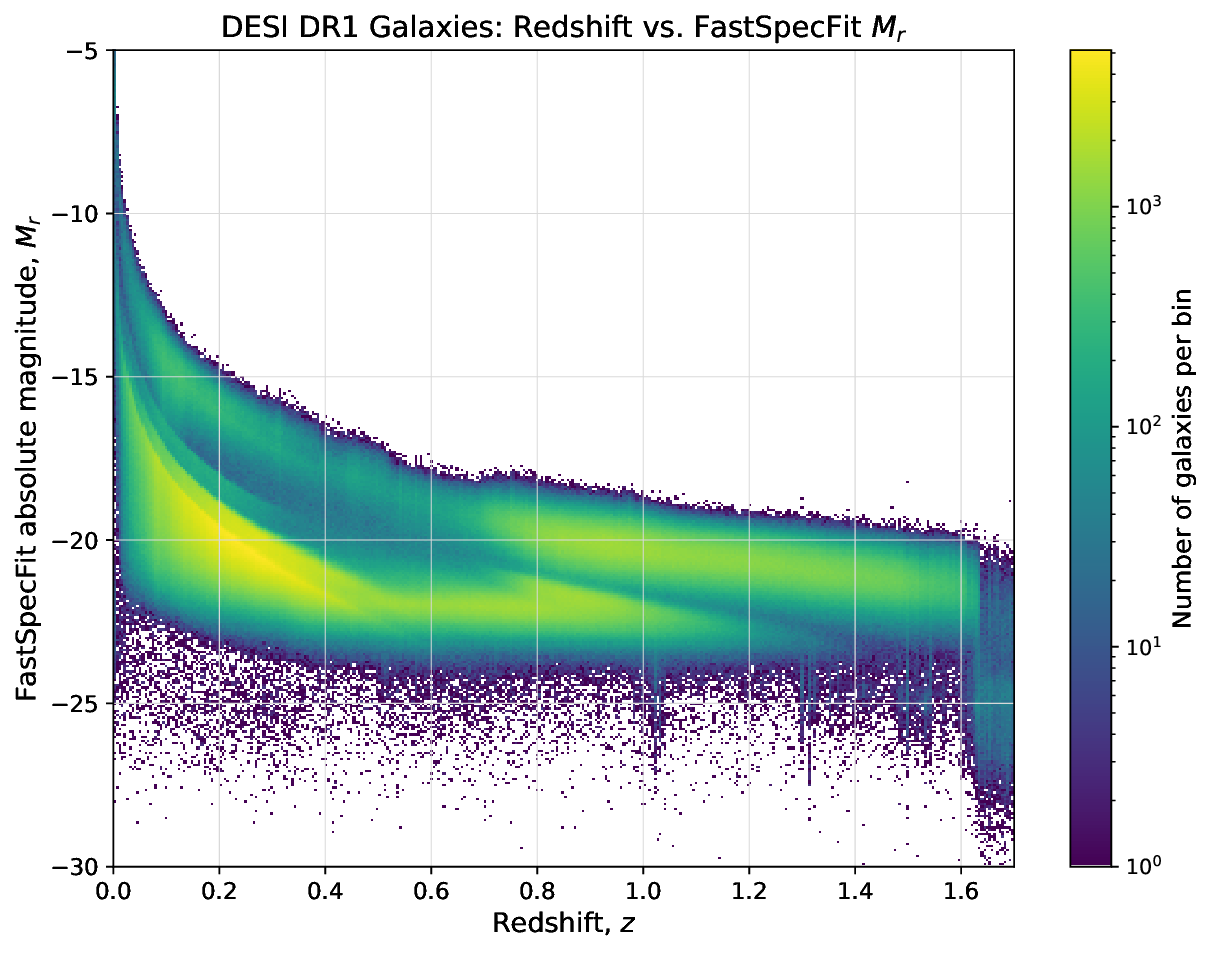}
            {0.4\textwidth}{(b)}
    }

    \gridline{
        \fig{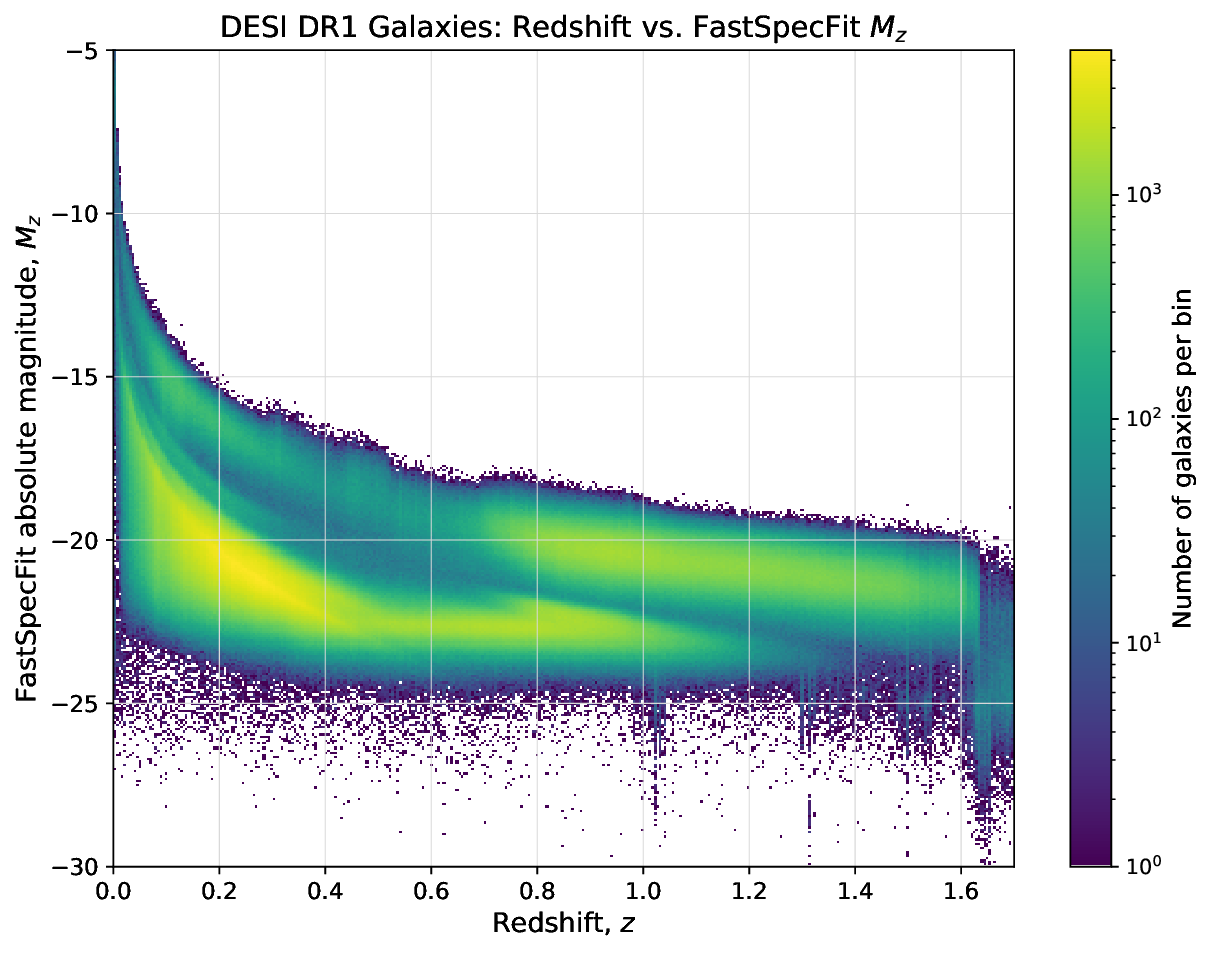}
            {0.4\textwidth}{(c)}
    }

    \caption{Rest-frame total absolute magnitude ($M_{\rm FastSpecFit}$) 
    versus redshift for the DESI DR1 sample across the 
    $g$, $r$, and $z$ bands. Magnitudes are taken directly from the 
    \texttt{FastSpecFit} Value-Added Catalog and include Galactic extinction 
    corrections, spectrophotometric K-corrections derived from SED modeling, 
    and normalization to integrated broad-band photometry to mitigate 
    fiber-aperture biases. Panels (a), (b), and (c) correspond to the 
    $g$, $r$, and $z$ bands, respectively.}

    \label{fig:mosaic_main}
\end{figure}
%

\section{Practical Prescriptions for Mitigating Systematic Biases} \label{sec:mitigation}

This section presents our practical framework: a unified, modular framework for correcting the dominant observational biases affecting DESI DR1. Figure~\ref{fig:mosaic_main} illustrates the distribution of DESI DR1 galaxies in the $z$-$M$ plane across the $g$, $r$, and $z$ bands. The observed structure highlights the strong coupling between target selection, luminosity-dependent incompleteness, and cosmological observation effects. To guide the reader through Figure~\ref{fig:mosaic_main}, this coupling is manifested across the $z$--$M$ distribution—where lighter color tones represent higher object density—through three prominent structural features (see also sections \ref{sec:correction}-\ref{sec:SBcorrection}): 
(i)~the sharp diagonal upper boundary in magnitude space, which traces the survey's apparent magnitude detection threshold driven by Malmquist bias; 
(ii)~the pronounced shifts in population density where the density peak (highlighted by the lighter color regions) jumps from intermediate-luminosity BGS targets to the highly luminous LRG sequence and active ELGs, directly reflecting the operational transitions and color-selection cuts (e.g., $M_g - M_r$) between target classes; and 
(iii)~the progressive narrowing of the magnitude distribution toward $z > 1.0$, where cosmological $(1+z)^{-4}$ Tolman surface-brightness dimming restricts the detectable sample exclusively to exceptionally bright systems. While unseen directly in the visual density map, Eddington bias further operates along these selection boundaries by scattering lower-luminosity objects into the detected sample via photometric uncertainties.


The following subsections summarize a flexible framework aimed at mitigating the dominant observational effects affecting both photometric and spectroscopic surveys. Depending on the scientific goals, individual components may be implemented separately or combined into a more comprehensive analysis. This flexible approach allows researchers to implement only the corrections relevant to their specific scientific objectives.

The goal of this section is not to prescribe a unique correction scheme, but rather to discuss the dominant observational effects that should be considered when interpreting the DESI DR1 fork geometry and when constructing subsamples intended for galaxy-evolution studies.

\subsection{Aperture, Extinction, and Rest-Frame Corrections}

\label{sec:correction}

The photometric measurements across redshift and morphology can be homogenized through the adoption of a unified correction formalism for the absolute magnitude ($M_{\rm corr}$):

\begin{equation}
\label{eq:Mcorr}
\begin{aligned}
M_{\rm corr} =\;& m_{\rm fib}
- 5\log_{10}\!\left[\frac{d_L(z)}{10\,{\rm pc}}\right]
- A_{\rm MW}(\lambda) - K(z) \\
&- \Delta M_{\rm ap}(\omega,n,z,r_e) + e(z) - \Delta M_{\rm Edd}(z) \\
&- A_{\rm int}(i,\lambda),
\end{aligned}
\end{equation}

where $m_{\rm fib}$ is the observed fiber magnitude, $d_L(z)$ is the luminosity distance, $A_{\rm MW}(\lambda)$ is the Galactic foreground extinction, $K(z)$ is the $K$-correction, $\Delta M_{\rm ap}$ accounts for aperture losses associated with the seeing ($\omega$), S\'ersic light profile ($n,z$) and color gradients ($r_e,z$), $e(z)$ is the evolutionary correction, $\Delta M_{\rm Edd}(z)$ is the Eddington bias correction, and $A_{\rm int}(i,\lambda)$ represents the inclination-dependent internal extinction.

Some of these corrections are physically straightforward and essential to establish a baseline rest-frame framework (e.g., $A_{\mathrm{MW}}$, $K(z)$). Conversely, others are heavily model-dependent, introducing external theoretical coupling and structural degeneracies that must be approached with caution (e.g., $e(z)$, $\Delta M_{\mathrm{Edd}}$, $A_{\mathrm{int}}$). For the diagnostic scope of this paper, these latter model-dependent terms are deliberately left unapplied; as demonstrated in Section~\ref{sec:lf_analysis}, maintaining these parameters uncorrected provides the empirical leverage necessary to isolate the unmitigated signatures of target-class mixing and survey selection boundaries. Below, we discuss the physical rationale and systematic impact of each individual term in detail.

\paragraph{Galactic Extinction ($A_{\rm MW}$):} Galactic extinction represents the dimming and reddening of extragalactic light caused by interstellar dust within our own Milky Way. This effect preferentially scatters shorter wavelengths, distorting the observed shape of the SED. Observed magnitudes are commonly corrected by subtracting the extinction term $A_{\rm MW}(\lambda)$ along the line of sight. This correction could be anchored to the \citet{Schlegel1998} dust maps, implementing the updated reddening calibration from \citet{Schlafly2011} and the standard wavelength-dependent extinction law characterized by \citet{Fitzpatrick1999}:
\begin{equation}
A_{\rm MW}(\lambda) = R_{\lambda} \cdot E(B-V),
\end{equation}
where $E(B-V)$ is the color excess and $R_{\lambda}$ is the specific extinction coefficient for each filter. Failure to apply this correction introduces systematic photometric offsets that artificially shifts galaxies toward the redder regions of the Green Valley, contaminating population statistics. For users of DESI DR1, we note that the \texttt{FastSpecFit} Value-Added Catalog \citep{Moustakas2023} already implements this multi-component correction self-consistently during its spectral and photometric scaling pipelines.

\paragraph{$K$-correction:} For the $K$-correction, we discuss a general framework in which the correction is preferably derived from robust spectrophotometric or multi-band photometric modeling rather than simplified single-template approximations. In any cosmological survey, since the observed data represent the source flux captured within a specific physical or instrumental window, the $K$-correction for a given band $R$ must be analytically defined by comparing the photon-weighted rest-frame flux to its redshifted counterpart through the same filter profile. Following the general K-correction formalism of \citet{Hogg2002} and the synthetic-photometry framework of \citet{Blanton2007}, the K-correction for a photon-counting detector can be expressed as:
\begin{equation}\label{eq:Kcorr_spectro}
K_R(z) = 2.5 \log_{10}(1+z) + 2.5 \log_{10} \left[ \frac{\int f_{\lambda, \rm rest}(\lambda) R(\lambda) \lambda d\lambda}{\int f_{\lambda, \rm rest}(\frac{\lambda}{1+z}) R(\lambda) \lambda d\lambda} \right],
\end{equation}
where $R(\lambda)$ is the filter transmission curve, and $f_{\lambda, \rm rest}$ is the rest-frame continuum of the galaxy. The inclusion of the $\lambda$ factor inside the integrands is mathematically required to correctly account for the photon-counting nature of modern CCD detectors. Within this unified framework, $f_{\lambda, \rm rest}$ can be reconstructed either from deep broad-band photometry via continuous SED fitting or directly from high-resolution spectroscopy. This modeling approach is inherently more robust than traditional empirical color-index shifts, as it captures how individual stellar population features—such as strong nebular emission lines or the prominent 4000 \AA\ break—shift across the filter windows as a function of cosmic time. For the specific case of DESI DR1 \citep{DESIDR1}, we suggest that researchers exploit the synergy of both regimes by utilizing the pre-computed $K$-corrections provided within the \texttt{FastSpecFit} Value-Added Catalog \citep{Moustakas2023}. Rather than performing a direct numerical integration over the raw, pixelated observational spectra—which is highly susceptible to noise inflation, cosmic ray artifacts, and camera gap losses—the \texttt{FastSpecFit} pipeline executes Equation~(\ref{eq:Kcorr_spectro}) by projecting the filter curves onto a noiseless, best-fit spectrophotometric model continuum that self-consistently pairs the broad-band imaging with the central spectroscopic flux. Users should verify that these profiles have been properly flux-calibrated to the total photometry scale using the pipeline's internal weights, helping minimize aperture-dependent systematics before drawing evolutionary conclusions.

\paragraph{Aperture Correction ($\Delta M_{\rm ap}$):} For the aperture correction, we consider it advantageous to account for missing flux due to atmospheric seeing and instrumental scaling, particularly when color gradients are expected to introduce redshift-dependent structural effects. Because of these gradients, the observed effective radius ($r_e$) systematically varies with the rest-frame wavelength being probed. Following the framework validated by \citet{vanderWel2014}, we suggest transforming all observed radii to a consistent rest-frame baseline using relations of the form $r_{e,{\rm rest}} = r_{e,{\rm obs}} / (1+z)^{\alpha}$, where $\alpha$ is the wavelength-dependent exponent derived from structural evolution models. This correction ensures that the spatial extrapolation of light profiles—typically modeled via S\'ersic extensions \citep{Graham2005}—helps minimize redshift-dependent structural biases, avoiding the artificial underestimation of luminosities at higher redshifts. These geometric corrections may become particularly important for nearby, spatially extended systems where a fixed physical aperture subtends only the innermost, bulge-dominated stellar regions. For the specific application to DESI DR1, however, we note that this explicit fiber aperture correction is substantially mitigated when utilizing the \texttt{FastSpecFit} Value-Added Catalog \citep{Moustakas2023}. The \texttt{FastSpecFit} pipeline inherently bypasses local fiber losses (arising from the $1.5''$ fiber diameter) by scaling the spectroscopic flux directly to the total, integrated broad-band photometry from the Legacy Surveys. Since these underlying imaging measurements are already derived from sophisticated, multi-band profile-fitting photometry optimized to recover near-total galaxy fluxes, the parameters are self-consistently normalized to the total emission scale, thereby reducing aperture-related selection systematics prior to sample definition.

\paragraph{Evolutionary Correction:} The evolutionary correction $e(z)$ attempts to account for the intrinsic luminosity evolution of stellar populations across cosmic time. In principle, galaxies at higher redshifts are expected to appear intrinsically brighter due to their younger stellar populations and more recent star-formation activity. A commonly adopted approximation parameterizes this effect as a linear relation, $e(z)=Q\cdot z$, where $Q$ is derived from Stellar Population Synthesis (SPS) models \citep{Bundy2006}. However, the precise form and normalization of this correction remain highly uncertain and strongly dependent on galaxy morphology, stellar mass, star-formation history, metallicity, and dust content. Consequently, evolutionary corrections are among the most model-dependent and least universally defined adjustments in observational galaxy studies. In heterogeneous galaxy populations, an improperly calibrated $e(z)$ term may artificially suppress or enhance apparent evolutionary trends, potentially introducing stronger systematic biases than those it attempts to mitigate. For this reason, we suggest treating evolutionary corrections with caution and applying them only when the adopted stellar population model is fully consistent with the scientific objectives of the analysis.

\paragraph{Eddington Bias Correction:} $\Delta M_{\rm Edd}(z)$ represents a statistical correction intended to mitigate the general Eddington bias, which arises when random observational uncertainties are convolved with a non-uniform underlying population density \citep{Eddington1913}. In galaxy surveys, this effect can systematically distort the observed luminosity distribution near detection boundaries by preferentially scattering the more numerous faint sources into intrinsically rarer bright bins \citep{Teerikorpi2004}. Following the analytical framework proposed by \citet{Fu2025}, a first-order approximation to this correction may be expressed as:
\begin{equation}
\Delta M_{\rm Edd}(z) = \alpha(z)\,\sigma_M^2(z)\,\ln(10),
\end{equation}
where $\sigma_M(z)$ represents the uncertainty in the absolute magnitude at a given redshift, and $\alpha(z)$ is the local logarithmic slope of the galaxy luminosity function evaluated near the target magnitude. Specifically, $\alpha(z)$ is defined as
\begin{equation}
\alpha (z) = \frac{d\log_{10}\phi(M,z)}{dM},
\end{equation}
where $\phi(M,z)$ denotes the galaxy luminosity function at absolute magnitude $M$ and redshift $z$. Larger values of $\alpha(z)$ correspond to regions where the luminosity function changes more rapidly with magnitude and are therefore more susceptible to Eddington bias. In principle, this formalism attempts to statistically compensate for the asymmetric migration of sources across luminosity bins induced by observational noise.

However, implementing this correction in practice remains highly non-trivial. The magnitude of $\Delta M_{\rm Edd}$ depends sensitively on the local shape of the luminosity function, which itself is strongly affected by sample incompleteness, surface-brightness selection effects, target-class mixing, and cosmic evolution. Furthermore, the correction becomes intrinsically coupled to the covariance between photometric uncertainties, structural measurements, and redshift-dependent selection functions. In heterogeneous surveys such as DESI, where multiple targeting strategies overlap across different galaxy populations, accurately determining the appropriate local luminosity-function slope is particularly challenging.

At low redshifts ($z \lesssim 0.3$), where photometric uncertainties remain relatively small, the net Eddington correction is expected to be minor compared to the dominant observational systematics. At higher redshifts, however, increasing photometric noise and cosmological surface-brightness dimming can amplify noise-driven population mixing near the survey boundaries. While applying $\Delta M_{\rm Edd}$ may help stabilize luminosity distributions in principle, an improperly calibrated correction can itself introduce artificial distortions by overcompensating the intrinsic population gradients. For this reason, we suggest treating Eddington corrections cautiously and primarily within analyses where the underlying completeness function and luminosity-function parametrization are independently well constrained.

\paragraph{Inclination-Dependent Internal Extinction ($A_{\rm int}$):} Inclination-driven internal extinction can substantially bias the observed photometric properties of highly inclined Late-Type Galaxies (LTGs; \citealt{Masters2010}). Consequently, dusty star-forming disks may be artificially displaced toward the Green Valley or passive sequence if geometric attenuation is neglected. The combined effects of internal extinction, surface-brightness dimming, and aperture-dependent light sampling can reach several tenths of a magnitude for highly inclined LTGs, especially over the range $0.1<z<1.0$ where extended late-type systems remain detectable and their structural measurements are highly sensitive to observational resolution effects before giving way to compact, point-like morphologies at higher redshifts.

A commonly adopted first-order correction models the inclination-dependent attenuation as \citep{Masters2010}:
\begin{equation}
A_{\rm int} = \gamma_{\lambda} \log_{10}(a/b),
\end{equation}
where $a/b$ is the observed semi-major to semi-minor axis ratio and $\gamma_{\lambda}$ is the band-dependent attenuation coefficient. In principle, this correction attempts to transform the observed luminosity to an approximate face-on orientation. However, the precise normalization of $\gamma_{\lambda}$ depends strongly on galaxy morphology, stellar mass, dust geometry, star-formation activity, and wavelength, rendering the correction intrinsically model-dependent and difficult to generalize across heterogeneous galaxy populations.

Consequently, applying an aggressive inclination-dependent attenuation correction may itself introduce systematic uncertainties comparable to those it attempts to mitigate. As an alternative, a more conservative strategy consists of limiting the analysis to galaxies whose orientations are less susceptible to severe projection and dust-attenuation effects. Such an approach reduces the dependence of the results on uncertain assumptions regarding dust geometry, stellar-population distributions, and the wavelength dependence of internal extinction.

An empirical motivation for this strategy is provided by \citet{Nigoche2026}, who found that dynamical-mass estimates become progressively less reliable for galaxies with inclination angles exceeding approximately $66^\circ$. Although this threshold was derived in the context of dynamical analyses, it is broadly consistent with previous observational studies reporting a rapid increase in projection, extinction, and structural uncertainties for highly inclined systems ($i \gtrsim 70^\circ$). Consequently, a conservative subsample defined by $i \lesssim 66^\circ$ may provide a useful compromise between sample completeness and robustness against orientation-driven systematics. While this selection does not completely eliminate internal attenuation, it substantially reduces the influence of heavily obscured edge-on galaxies and mitigates one of the principal sources of inclination-dependent scatter in luminosity-based analyses. By restricting the contribution of extreme viewing geometries, the resulting galaxy sample is expected to provide a more homogeneous basis for investigating intrinsic luminosity distributions and their evolution across cosmic time.

\subsection{Kinematic Aperture Normalization}

Fiber-integrated velocity dispersions ($\sigma_{\rm obs}$) are intrinsically affected by aperture-dependent sampling effects. Because stellar velocity-dispersion profiles generally decrease with galactocentric radius, a fixed angular aperture probes different physical regions of a galaxy as a function of redshift. In the DESI framework, the $1.5''$ fiber preferentially samples the dynamically hotter central regions of nearby galaxies, while progressively integrating larger and kinematically colder regions at higher redshifts. If left uncorrected, this effect can introduce systematic distortions in dynamical scaling relations and bias dynamical-mass estimates derived from relations of the form $M_{\rm dyn}\propto\sigma^2 r_e$.

A commonly adopted first-order correction rescales the observed velocity dispersion to a standardized physical aperture, typically defined relative to the effective radius \citep{Jorgensen1995}:
\begin{equation}
\sigma_{\rm corr} = \sigma_{\rm obs} \left( \frac{R_{\rm fiber}}{R_{\rm norm}} \right)^{\delta},
\end{equation}
in DESI observations $R_{\rm fiber}=0.75''$ represents the fiber radius, $R_{\rm norm}$ is the reference aperture (often normalized to $r_e/8$), and $\delta$ parameterizes the slope of the velocity-dispersion profile. Empirical studies based on integral-field spectroscopy suggest characteristic values of $\delta\approx-0.066$ for early-type galaxies \citep{Cappellari2006}, while somewhat steeper profiles have been reported for late-type systems.

In principle, this normalization attempts to place galaxies observed at different redshifts onto a common dynamical scale. However, implementing the correction robustly remains non-trivial. The slope $\delta$ is not universal and depends on morphology, stellar mass, rotational support, bulge fraction, inclination, environment, and the presence of kinematically distinct substructures. Furthermore, the correction becomes increasingly uncertain for late-type galaxies, where rotational broadening, beam smearing, and asymmetric line profiles complicate the interpretation of the integrated fiber dispersion.

Additional uncertainties arise from the coupling between seeing conditions, effective-radius measurements, and the physical scale subtended by the fiber at different redshifts. In heterogeneous surveys such as DESI, where multiple galaxy populations coexist across a broad range of structural properties, adopting a single universal aperture correction may therefore introduce systematic effects comparable to those it seeks to mitigate.

At low redshift, where galaxies are well resolved and the fiber samples only the innermost regions, aperture effects can become significant and may bias direct comparisons between nearby and distant systems. At intermediate and higher redshifts, however, the larger physical coverage of the fiber partially reduces these gradients for compact galaxies, although residual systematics remain. Consequently, while aperture-normalized velocity dispersions may improve the consistency of dynamical analyses in principle, we suggest applying these corrections cautiously and only when the adopted structural parametrization and kinematic assumptions are fully compatible with the scientific objectives of the study.

\subsection{Malmquist Bias and Completeness Limits}

\label{subsec:aperture_extinction_restframe}

The dominant selection effect in the $z$--$M$ plane is the Malmquist bias, which artificially shifts the observed population toward intrinsically brighter galaxies at increasing redshift. To preserve statistical completeness, a useful strategy consists of restricting the analysis to a volume-limited region satisfying:
\begin{equation}
M_{\rm corr} < M_{\rm comp}(z),
\end{equation}
where $M_{\rm comp}(z)$ is obtained by evaluating Equation~(\ref{eq:Mcorr}) at the survey limiting magnitude $m_{\rm lim}$ or at some chosen magnitude, depending on the needs of the research. This definition helps ensure internal consistency by avoiding the double application of distance modulus, $K$-corrections, and evolutionary terms already included in Equation~(\ref{eq:Mcorr}).

The completeness region should be constructed through the following procedure:
(1) define the redshift interval $[z_{\rm min},z_{\rm max}]$;
(2) compute $M_{\rm comp}(z)$ at $z_{\rm max}$ using Equation~(\ref{eq:Mcorr}) evaluated at $m_{\rm lim}$;
(3) retain only galaxies satisfying $M_{\rm corr}<M_{\rm comp}(z)$ throughout the selected redshift interval, ensuring the sample is restricted to the regions outside the observational gaps inherent to the survey's targeting strategy.

By adopting this strategy, selection biases can be reduced, facilitating more homogeneous comparisons between nearby and distant populations. Within this region, the use of quasi-constant magnitude (mass) bins ($\Delta \log M = 0.1$ dex) and narrow redshift intervals ($\Delta z = 0.01$) is suggested to track dynamical and structural properties while minimizing observational and instrumental interference (e.g., \citealt{Nigoche2026}).

\subsection{Surface-Brightness Corrections and Volume Completeness}
\label{sec:SBcorrection}

We additionally emphasize the need to account for cosmological surface-brightness dimming (Tolman dimming), which scales as $(1+z)^{-4}$. While this effect is analytically included in the definition of the luminosity distance ($d_L$) considered in $M_{\rm corr}$, its impact on the limiting surface-brightness ($\mu_{\rm lim}$) may contribute to the observed depletion of galaxies in the Green Valley region at $z \gtrsim 0.7$ by pushing extended, low-surface-brightness sources below the detection threshold. Given that DESI spectroscopy relies on a fixed $1.5''$ fiber aperture, diffuse outer regions contribute progressively less flux at increasing redshift, particularly for extended LTGs. Therefore, it is important to recognize that this effect biases luminosity, stellar-mass, and structural measurements by preferentially excluding low-surface-brightness systems. This loss of completeness is further compounded by the general Eddington bias. The interplay between the cosmological disappearance of extended sources and noise-driven photometric scattering may contribute to an artificial blending of the Quenched and Star-Forming branches. To mitigate these combined biases and ensure a reliable comparison across cosmic time, the effective surface-brightness must first be purged of external dimming and obscuration effects to isolate the intrinsic density of the sources. We define the rest-frame corrected effective surface-brightness ($\mu_{\rm corr}$) as:
\begin{equation}
\label{eq:mucorr}
\begin{aligned}
\mu_{\rm corr} =\;& \mu_{\rm obs} - 10\log_{10}(1+z) - A_{\rm MW}(\lambda) - K(z) \\
&- A_{\rm int}(i,\lambda) + e(z),
\end{aligned}
\end{equation}
where $A_{\rm MW}(\lambda)$ is the Galactic foreground extinction, $K(z)$ is the $K$-correction, $A_{\rm int}(i,\lambda)$ represents the inclination-dependent internal dust extinction, $e(z)$ is the evolutionary correction, and $\mu_{\rm obs}$ is the raw, observed effective surface-brightness. This latter parameter is calculated within the effective radius $r_e$ (expressed in arcseconds) from the apparent magnitude $m$ via:
\begin{equation}\label{eq:muobs}
\mu_{\rm obs} = m + 2.5\log_{10}(2\pi r_e^2).
\end{equation}
The term $10\log_{10}(1+z)$ in Equation~(\ref{eq:mucorr}) explicitly compensates for the cosmological Tolman surface-brightness dimming.

We caution against calculating the effective surface-brightness using the flux captured within the spectroscopic fiber aperture ($m_{\rm fib}$). Because fibers probe a fixed angular size ($1.5''$ for DESI), utilizing fiber magnitudes would introduce a severe, distance-dependent structural bias that distorts the coupling between spatial profiles and cosmic time. Instead, $\mu_{\rm obs}$ is preferably derived from total broad-band imaging photometry. By modeling the overall light distribution of the galaxy—via full-profile S\'ersic fits or curve-of-growth analyses—broad-band photometry provides a robust, aperture-independent measurement of the total apparent magnitude $m$ and the true structural half-light radius $r_e$. This distinction helps to safeguard the structural integrity of the sample, as it prevents local aperture losses from being misidentified as intrinsic cosmological evolution or artificial quenching in the high-redshift regime.

We note that a formal multidimensional Eddington correction ($\Delta M_{\rm Edd}$) for surface-brightness is intrinsically coupled to the covariance between flux and size measurements. To mitigate these coupled uncertainties, we instead adopt a conservative volume-completeness threshold. Unlike the absolute-magnitude formalism of Equation~(\ref{eq:Mcorr}), the surface-brightness relation in Equation~(\ref{eq:mucorr}) does not explicitly include an aperture-correction term. Surface-brightness, as an intensive quantity defined per unit angular area, is generally less sensitive to global aperture losses than integrated luminosity measurements, although residual profile-dependent effects may remain for centrally concentrated systems. Nevertheless, all measurements remain subject to the cosmological Tolman dimming term, $(1+z)^{-4}$, which dominates the redshift evolution of the observed surface-brightness distribution. Under these conditions, the adopted completeness limits provide a statistically controlled framework for minimizing surface-brightness selection biases.

We suggest restricting the scientific sample across all epochs to galaxies satisfying a strict, rest-frame structural integrity criterion applied to this corrected parameter:
\begin{equation}
\label{eq:mulim}
\mu_{\rm obs} < \mu_{\rm lim},
\end{equation}
where $\mu_{\rm lim}$ is a fixed, observed-frame surface-brightness threshold. This limit should be determined empirically from the survey completeness function at the highest redshift bin of interest by identifying the regime where galaxy counts begin to decline relative to the expected distribution within narrow redshift bins due to sky-noise dominance. By truncating both low- and high-redshift sub-samples at this same intrinsic boundary, we substantially reduce structural selection biases. This integrated approach may help stabilize the location of the Green Valley boundary and improve the likelihood that the observed fork geometry reflects genuine quenching signatures rather than observational scatter or target-selection effects.

\begin{figure}[ht]
    \centering
    \includegraphics[width=0.4\textwidth]{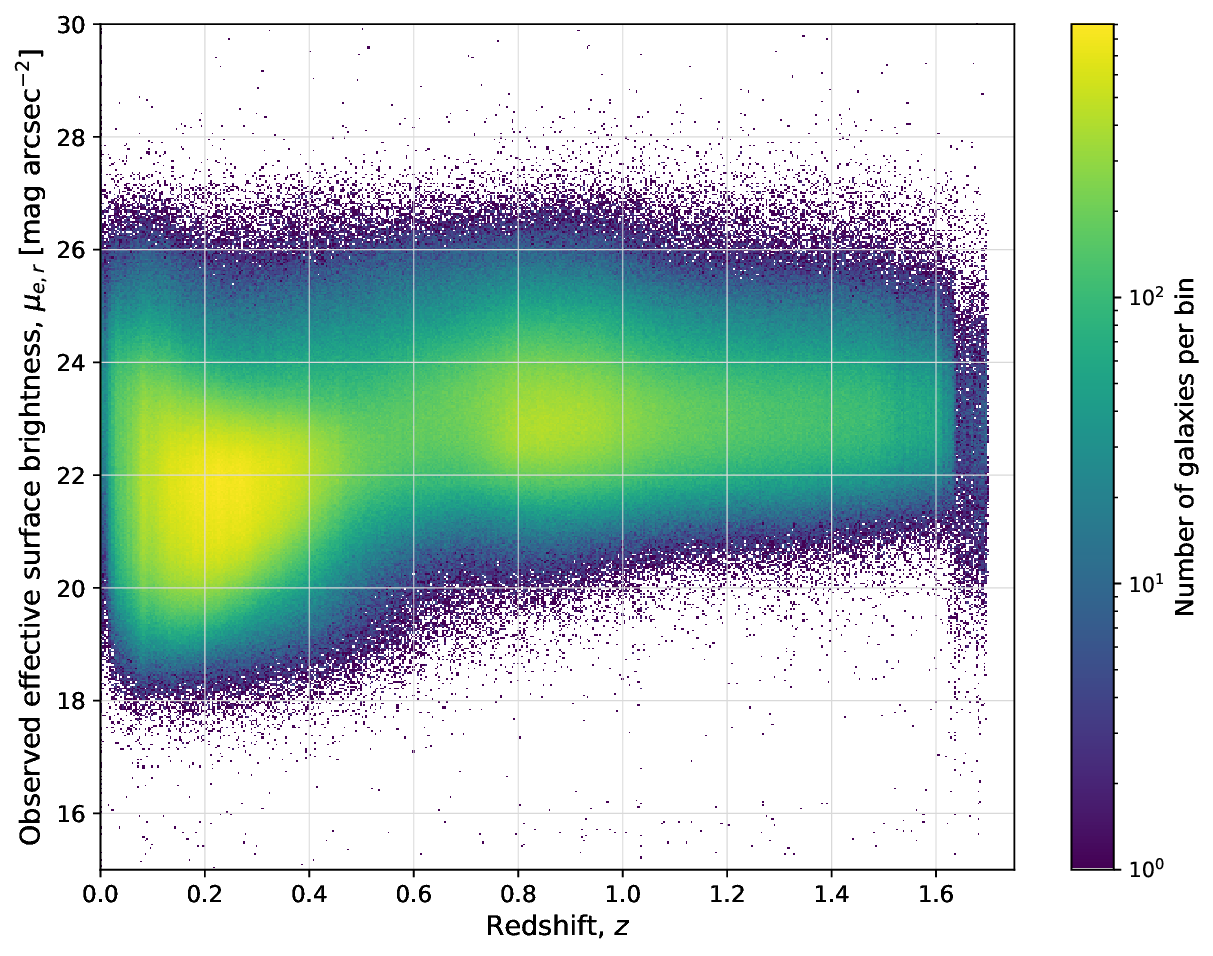}
    \caption{Observed effective surface-brightness in the $r$ band, $\mu_{\mathrm{eff},r}$, as a function of redshift for DESI DR1 galaxies. The effective surface-brightness was computed from the Legacy Surveys apparent magnitude and effective radius as $\mu_{\mathrm{eff},r}=m_r+2.5\log_{10}\left(2\pi R_{\mathrm{eff}}^2\right)$, using \texttt{mag\_r} and \texttt{shape\_r}. The density map highlights the main galaxy population and the surface-brightness selection structure across redshift.}
    \label{fig:z_vs_mu}
\end{figure}

\begin{figure}[ht]
    \centering
    \includegraphics[width=0.4\textwidth]{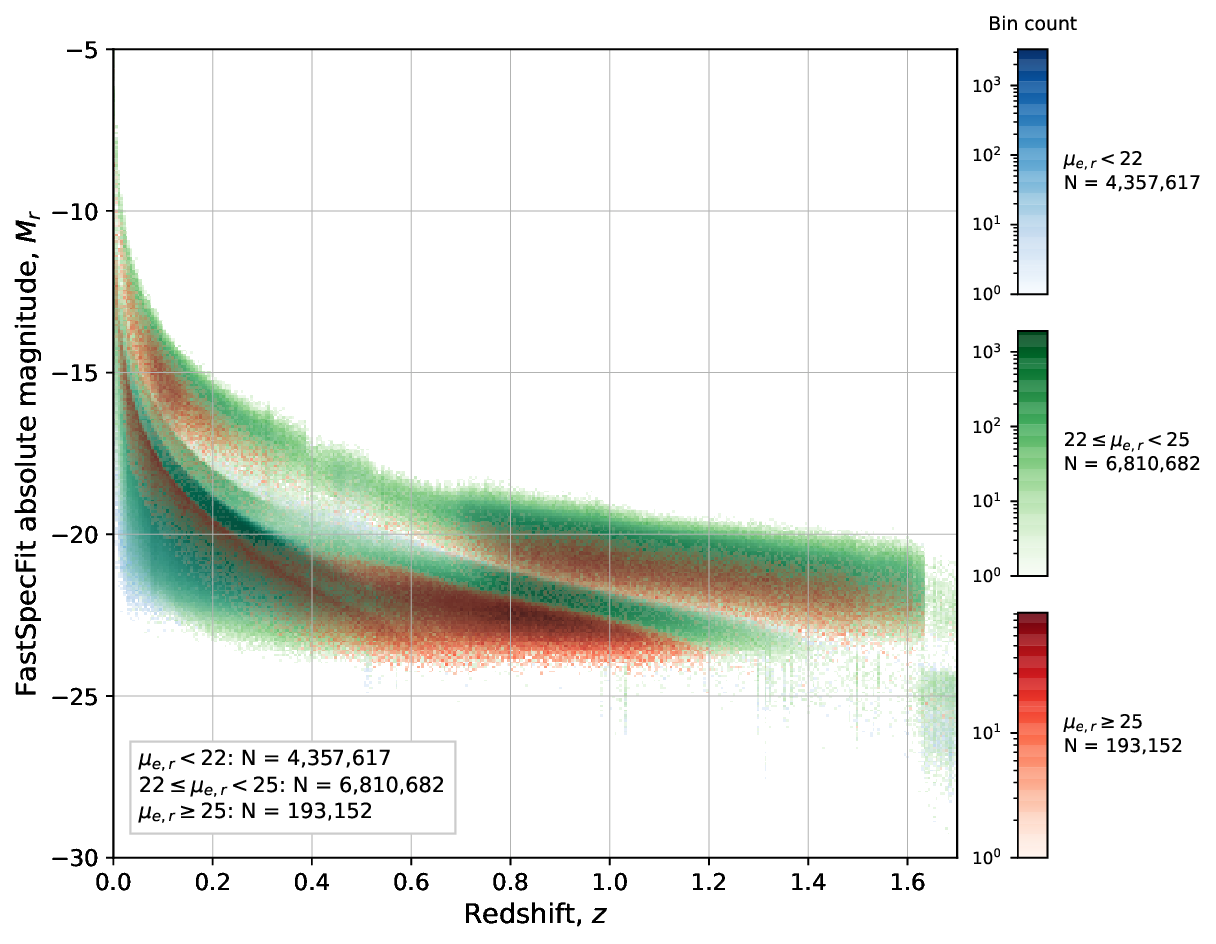}
    \caption{Distribution of DESI DR1 galaxies in the redshift--absolute magnitude plane in the $r$ band ($z$ vs. $M_{\mathrm{FastSpecFit}}$), color-coded by observed effective surface-brightness $\mu_{\mathrm{eff},r}$ in three surface-brightness regimes: blue for $\mu_{\mathrm{eff},r} < 22$ mag arcsec$^{-2}$ (high surface-brightness), green for $22 \leq \mu_{\mathrm{eff},r} \leq 25$ mag arcsec$^{-2}$ (intermediate surface-brightness), and red for $\mu_{\mathrm{eff},r} > 25$ mag arcsec$^{-2}$ (low surface-brightness). The color coding illustrates how different surface-brightness populations occupy distinct regions of the redshift--absolute magnitude plane.}
    \label{fig:z_vs_M_overlay}
\end{figure}

\begin{figure}[ht]
    \centering

    \gridline{
        \fig{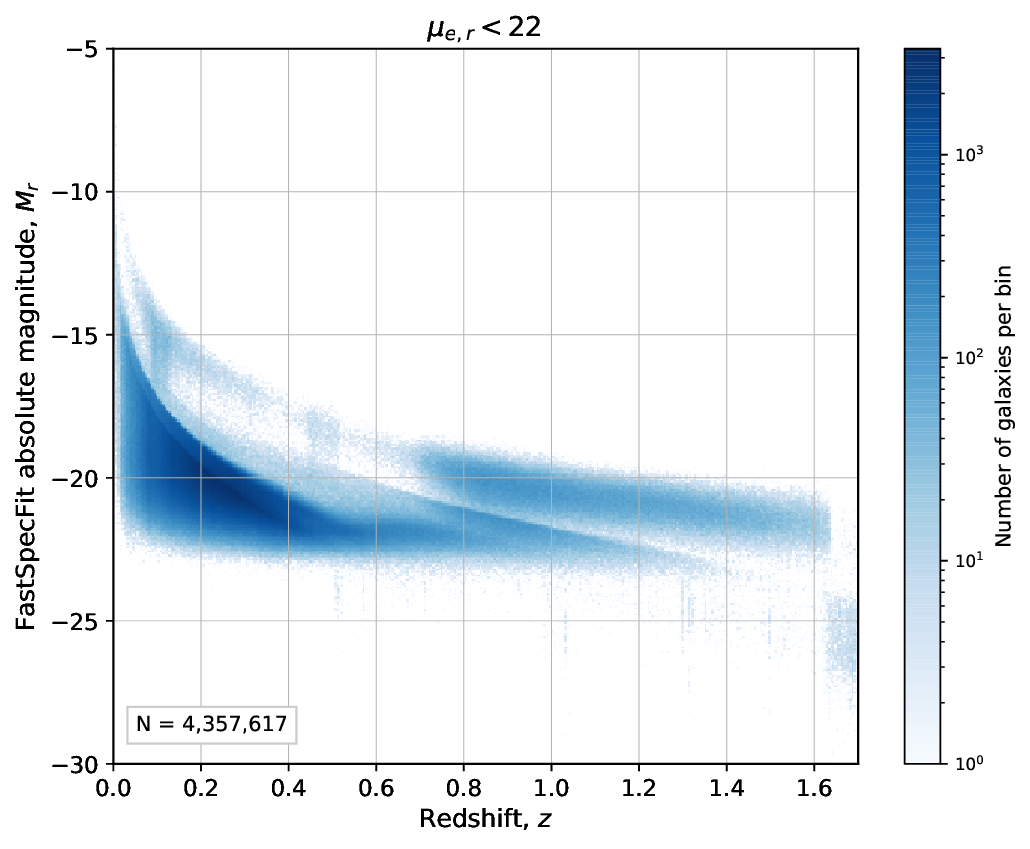}
            {0.4\textwidth}{(a)}
        \fig{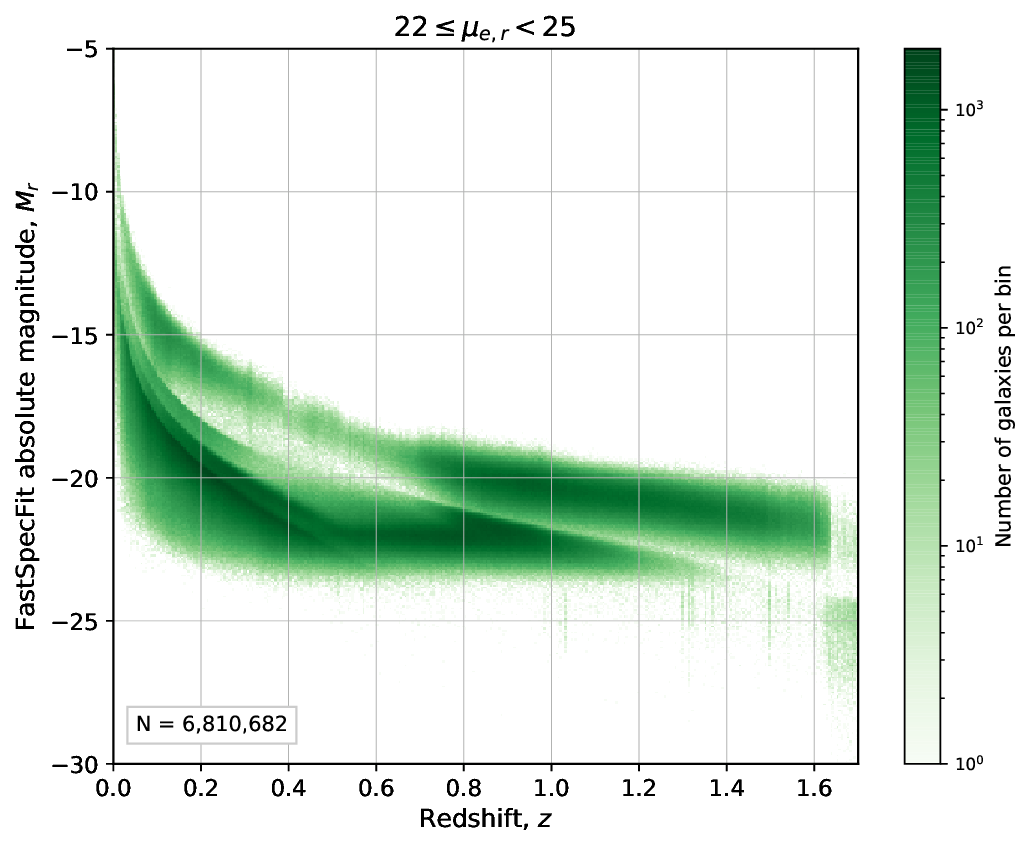}
            {0.4\textwidth}{(b)}
    }

    \gridline{
        \fig{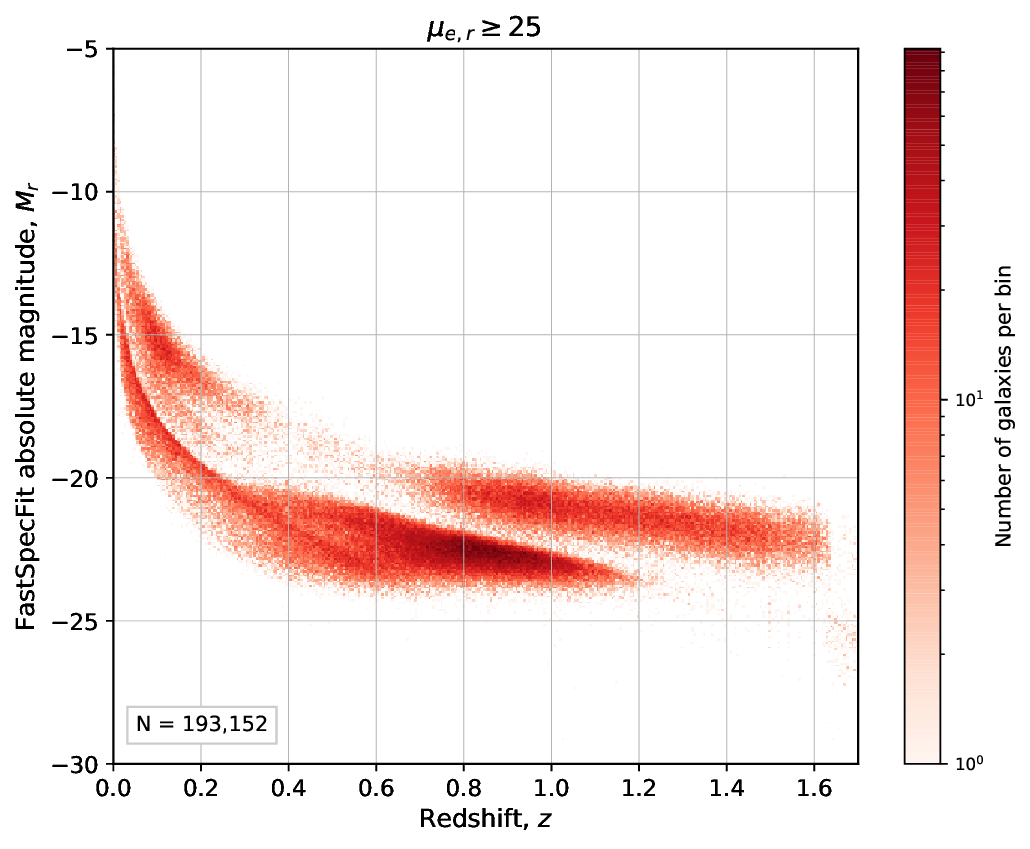}
            {0.4\textwidth}{(c)}
    }

    \caption{Distribution of DESI DR1 galaxies in the redshift--absolute
    magnitude plane in the $r$ band ($z$ vs. $M_{\rm FastSpecFit}$), with
    the sample divided according to three observed effective
    surface-brightness regimes. Panel (a) shows
    $\mu_{\mathrm{eff},r} < 22$ mag arcsec$^{-2}$
    (high surface-brightness), panel (b) shows
    $22 \leq \mu_{\mathrm{eff},r} \leq 25$ mag arcsec$^{-2}$
    (intermediate surface-brightness), and panel (c) shows
    $\mu_{\mathrm{eff},r} > 25$ mag arcsec$^{-2}$
    (low surface-brightness).}

    \label{fig:z_vs_M_panels}
\end{figure}

The relevance of these considerations is illustrated by the empirical distributions shown in Figures~\ref{fig:z_vs_mu}, \ref{fig:z_vs_M_overlay}, and \ref{fig:z_vs_M_panels}.

In Figure~\ref{fig:z_vs_mu}, we map redshift against the observed-frame surface-brightness in the $r$-band ($\mu_{\rm eff, \textit{r}}$). This distribution uncovers the severe impact of the cosmological Tolman dimming; as redshift increases, the locus of the population shifts systematically toward fainter surface-brightnesses. Concurrently, the sample selection function truncates abruptly at the faint end ($\mu_{\rm eff, \textit{r}} \gtrapprox 25$ mag arcsec$^{-2}$) as it encounters the sky-noise dominance regime, resulting in a drastic decrease in galaxy counts. This diagnostic identifies the regime where structural incompleteness becomes increasingly important, marking the approximate boundary where the detection probability $P(\mu)$ drops below unity.

Crucially, at the highest redshift edges of this isolated distribution, the observed behavior is consistent with the expected signatures of Eddington bias. As the signal-to-noise ratio degrades with distance, the scatter of the population visibly narrows. This behavior is a classic statistical artifact: near the survey's detection limit, photometric and structural uncertainties preferentially scatter intrinsically fainter galaxies into brighter categories, skewing the observed locus of the population. Enforcing the strict empirical threshold from Equation~\ref{eq:mulim} is therefore a useful step toward mitigating to filtering out both the cosmological dimming artifacts and these noise-driven Eddington scattering effects, allowing us to approach a substantially less biased representation of the underlying evolutionary sequence.

To trace how this structural selection propagates into the fundamental parameter space, Figure~\ref{fig:z_vs_M_overlay} presents redshift versus the rest-frame total absolute magnitude, color-stratified into three distinct surface-brightness regimes: high ($\mu_{\rm eff, \textit{r}} < 22$ mag arcsec$^{-2}$), intermediate ($22 \le \mu_{\rm eff, \textit{r}} \le 25$ mag arcsec$^{-2}$), and low ($\mu_{\rm eff, \textit{r}} > 25$ mag arcsec$^{-2}$) surface-brightness galaxies. Due to the intrinsic blending of multi-colored layers and the saturation of overlapping data points, a definitive interpretation of the underlying densities within the composite ``fork'' geometry remains challenging in this overlay view.

To overcome the visual limitations of the overlay and rigorously dissect the distribution of densities, the multi-panel mosaic in Figure~\ref{fig:z_vs_M_panels} isolates these three surface-brightness cohorts, revealing how structural selection maps onto the two main branches of the fork. In the high-surface-brightness panel ($\mu_{\rm eff, \textit{r}} < 22$ mag arcsec$^{-2}$), the highest density concentration is strictly confined to low redshifts ($z < 0.4$). As redshift increases, this cohort displays a highly asymmetric distribution between the two branches of the fork. Along the upper branch, the density is remarkably low at redshifts below $z \approx 0.7$; however, beyond this boundary, the population density experiences a noticeable increment, subsequently holding a semi-constant distribution that gradually declines as it approaches $z \approx 1.6$. Conversely, the lower branch maintains a prominent and highly evident density at low distances, which then progressively diminishes until completely fading away near $z \approx 1.4$.

Conversely, the intermediate-surface-brightness panel ($22 \le \mu_{\rm eff, \textit{r}} \le 25$ mag arcsec$^{-2}$) uncovers the most structurally complete, continuous, and widely extended distribution in the parameter space. Here, the galaxy population densely traces both the upper and lower branches of the fork simultaneously. This intermediate regime bridges the structural gap, showcasing the coeval development of both sequences before selection thresholds truncate the lower branch at higher redshifts. Finally, the low-surface-brightness panel ($\mu_{\rm eff, \textit{r}} > 25$ mag arcsec$^{-2}$) shows that the high-density locus does not emerge until $z > 0.4$, selectively embedding itself almost exclusively along the lower branch of the fork and stretching up to $z \approx 1.2$, where the entire cohort experiences a sharp volume truncation.

Furthermore, analyzing these isolated panels in tandem with Figure~\ref{fig:TargetClasseSMr} strongly suggests that the observed density configurations along both branches of the fork are primarily driven by the overlapping selection functions of the principal DESI target classes. At low redshifts ($z < 0.4$), the prominent local density core seen in the high-surface-brightness ($\mu_{\rm eff, \textit{r}} \le 22$) panel maps is broadly consistent with the Bright Galaxy Survey (BGS) selection function. The BGS also supplies the bulk of the intermediate-surface-brightness galaxies at these low distances. However, beyond $z = 0.4$, the survey experiences a drastic population transition. The continuous high-density structure observed along the lower branch of the fork across both the intermediate ($22 \le \mu_{\rm eff, \textit{r}} \le 25$)- and low-surface-brightness panels ($\mu_{\rm eff, \textit{r}} > 25$) is driven by the massive onset of the Luminous Red Galaxy (LRG) target class. These massive, passive spheroids possess high central stellar densities that keep them compact and structurally resilient against Tolman dimming. The sharp termination of this dense structure at $z \approx 1.2$ is consistent with the flux boundary where the LRG selection function reaches its volume limit.

Concurrently, the upper branch of the fork undergoes a complex systematic migration. At high surface-brightnesses ($\mu_{\rm eff, \textit{r}} \le 22$) this upper branch is initially populated by the Emission Line Galaxy (ELG) selection function, which captures star-forming disks. However, as cosmological dimming takes effect, the bulk of this ELG statistical density is artificially downshifted into the intermediate-surface-brightness ($22 \le \mu_{\rm eff, \textit{r}} \le 25$) and low-surface-brightness panel ($\mu_{\rm eff, \textit{r}} > 25$), creating the illusion of a sudden structural transformation along the upper branch. The sharp termination of this dense structure at $z \approx 1.6$ marks the effective flux boundary where the ELG selection function reaches its volume limit. Finally, the high-redshift tail extending up to $z \approx 1.6$ along the upper branch of the high-surface-brightness panel aligns also with the entry of the point-like Quasar (QSO) target class and extreme ELG outliers. Under low signal-to-noise conditions at high redshift, profile-fitting algorithms cannot resolve these quasar central emission from their host galaxies, artificially concentrating their integrated light into high surface-brightness values ($\mu < 22$). However, as visually confirmed by the multi-panel mosaic, this regime exhibits a very low, sparse density of sources compared to the dominant intermediate and low surface-brightness channels, highlighting that these high-redshift, ultra-compact targets remain clear statistical outliers within the survey framework. Mapping these target classes indicates that the apparent convergence of the fork at $z > 0.7$ is largely driven by population blending from massive LRGs to compact ELGs and point-like QSOs, rather than by a physical disappearance of galaxy bimodality. This motivates the use of our joint $\mu_{\rm lim}$ constraint to decouple true physical evolution from distance-dependent structural trends.

\section{Quantitative Impact on Galaxy Luminosity Function Parameters}
\label{sec:lf_analysis}

Having established the qualitative impact of surface-brightness selection on the fork geometry in Section~\ref{sec:SBcorrection}, we now turn to a quantitative evaluation using the galaxy luminosity function (LF). By fitting standard functional forms to raw, target-selected spectroscopic samples across $0.05 < z < 1.6$, we map how observational selection boundaries artificially deform inferred population parameters ($\phi^*, M^*, \alpha$). In this section, we outline the diagnostic rationale and limitations of these parametric fits before detailing the non-parametric estimator methodology.



\subsection{Limitations of the LF Analysis}
\label{sec:limitations}

Before applying estimator algorithms to the DESI DR1 dataset, it is critical to clarify the methodological purpose of the luminosity functions (LFs) derived in this work. Standard galaxy evolution studies require LFs constructed from volume-limited, completeness-corrected samples \citep[e.g.,][]{Blanton2003, Driver2005}. In contrast, the primary objective of this section is to utilize parametric fits as a quantitative diagnostic tool to evaluate spectroscopic target selection incompleteness \citep{Myers2023}. By fitting standard Schechter functions to uncorrected target-selected spectroscopic samples across the redshift range $0.05 < z < 1.6$, we explicitly map how the interplay of target-class boundaries, cosmological attenuation, flux-limit truncation, and instrumental constraints artificially distort inferred galaxy population parameters ($\phi^*, M^*, \alpha$). Furthermore, the parameter uncertainties derived from $1/V_{\mathrm{max}}$ Poisson statistics represent statistical lower limits that omit cosmic variance and systematic selection errors. Consequently, these evaluations do not constitute a completeness-corrected measurement of intrinsic galaxy populations, and their interpretation is subject to several key observational limitations:


\begin{enumerate}
    \item \textbf{Non-smooth selection function:} The $1/V_{\mathrm{max}}$ estimator assumes a smooth selection function, which is explicitly violated at the boundaries between target classes (BGS, LRG, ELG, QSO).

    \item \textbf{Fiber assignment and target priority} -- The $1/V_{\mathrm{max}}$ calculation assumes uniform spatial coverage. However, DESI's hierarchical fiber allocation introduces systematic completeness biases. In the dark-time program ($z > 0.4$), priority favors QSO and LRG targets over ELG candidates, dropping diffuse systems via fiber collisions in dense regions. Conversely, the low-redshift domain ($z < 0.4$) is governed by the bright-time program, where BGS Bright takes precedence over BGS Faint.

    \item \textbf{Geometric area normalisation} -- The absolute spatial density scale $\phi^*$ is currently uncorrected for the partial sky coverage of the DESI DR1 footprint. While omitting this global geometric factor shifts the overall vertical normalisation of the LF curves into arbitrary units, it leaves the intrinsic shapes, slopes ($\alpha$), and characteristic magnitudes ($M^*$) completely unaltered, thereby preserving the diagnostic validity of the population-blending analysis presented here.

    \item \textbf{Absence of cosmic luminosity evolution} -- No empirical or theoretical luminosity evolution correction, $e(z)$, was applied to the rest-frame absolute magnitudes. While omitting this correction leaves the sample exposed to the passive aging of stellar populations at high redshift—which artificially brightens intrinsic magnitudes and displaces the characteristic parameter $M^*$ in older systems—this omission was intentional to isolate purely instrumental selection effects from exogeneous evolutionary models, preserving the structural diagnostic framework of our analysis.

    \item \textbf{Absence of inclination corrections} -- Rest-frame absolute magnitudes were not corrected for internal dust extinction as a function of galactic inclination. While omitting this adjustment introduces systematic dimming in highly inclined disc systems, executing such structural corrections requires robust morphology-dependent axis-ratio measurements. For the diagnostic scope of this paper, omitting inclination parameters avoids introducing external structural model dependencies, ensuring that the detected discontinuities remain strictly tied to the primary DESI target-selection boundaries.
    
    \item \textbf{Absence of surface-brightness constraints:} Although total absolute magnitudes were corrected for aperture effects, galactic extinction, $K$-corrections, and luminosity distance, no surface-brightness selection cuts were enforced during the LF calculation. A robust correction requires the redshift-dependent transformation from Equations~(\ref{eq:mucorr}) and~(\ref{eq:mulim}).
    
    \item \textbf{No Eddington bias correction:} No explicit Eddington bias correction was applied, as its implementation requires a well-constrained luminosity function model that is itself biased by the same selection effects we seek to diagnose.

    \item \textbf{Methodological scope:} The LF analysis is designed to quantify and map the magnitude of the distortion rather than to eliminate it. Omitting a multi-dimensional boundary layer treatment leads directly to the systematic instabilities demonstrated in Figures~\ref{fig:phi_star},~\ref{fig:m_star}, and~\ref{fig:alpha}.


\end{enumerate}

Future work combining DESI DR1 data with forward simulations of the full selection function — including realistic mock catalogs and survey footprint models — will be required to quantify the remaining systematic uncertainties and to refine evolutionary measurements beyond the diagnostic level presented here.

The parametric fits presented herein are explicitly designed as visual and quantitative tools to map where selection completeness breaks down across target classes. The practical implications of these diagnostic results for cosmological pipelines—along with a structured framework for downstream analysis mitigations—are further detailed in Section~\ref{sec:mitigation_framework}.

\subsection{Methodology}\label{sec:methodology}

We reconstruct the differential galaxy space density in redshift bins of $\Delta z = 0.1$ using the non-parametric $1/V_{\rm max}$ estimator \citep{Schmidt1968}:

\begin{equation}
\label{eq:vmax_sum}
\phi(M) = \frac{1}{\Delta M} \sum_{i} \frac{1}{V_{\rm max, i}},
\end{equation}
with Poisson uncertainties:
\begin{equation}
\label{eq:vmax_err}
\sigma_{\phi} = \frac{1}{\Delta M} \sqrt{\sum_{i} \frac{1}{V_{\rm max, i}^2}}.
\end{equation}

where $V_{\mathrm{max}, i}$ represents the maximum comoving volume over which galaxy $i$ could remain within the survey's nominal flux limits. 

We note that the standard application of Equation~(\ref{eq:vmax_sum}) herein assumes uniform spatial coverage and full-sky geometry. In practice, the true spatial completeness of the DESI footprint is modulated by global geometric area restrictions and a hierarchical fiber-assignment priority scheme (which privileges specific target classes like QSOs and LRGs, inducing localized fiber collisions). For the diagnostic scope of this work, we deliberately omit these correction factors from the estimator. This allows us to map the unmitigated imprint of instrumental boundaries directly onto the resulting Schechter parameters, with a comprehensive breakdown of these systematic effects provided in Section ~\ref{sec:methodology}.

We fit the resulting LFs with the Schechter function \citep{Schechter1976}:
\begin{equation}
\label{eq:schechter}
\begin{aligned}
\phi(M) \, dM =\;& 0.4 \, \ln(10) \, \phi^* \, 10^{0.4(\alpha+1)(M^*-M)} \\
&\times \exp\!\left[-10^{0.4(M^*-M)}\right] dM,
\end{aligned}
\end{equation}
using non-linear least squares in $\log_{10}\phi$ space. Where $M^*$ represents the characteristic breaking magnitude, $\alpha$ is the faint-end slope, and $\phi^*$ is the normalization density. This procedure is applied to the full combined sample and independently to the four mutually exclusive target classes: BGS, LRG, ELG, and QSO. Mixed target assignments (e.g., BGS|LRG) are excluded.


\begin{figure*}[t]
    \centering

    \setlength{\tabcolsep}{2pt}

    \begin{tabular}{@{}cc@{}}

    \includegraphics[
        width=0.49\textwidth,
        height=0.195\textheight,
        keepaspectratio
    ]{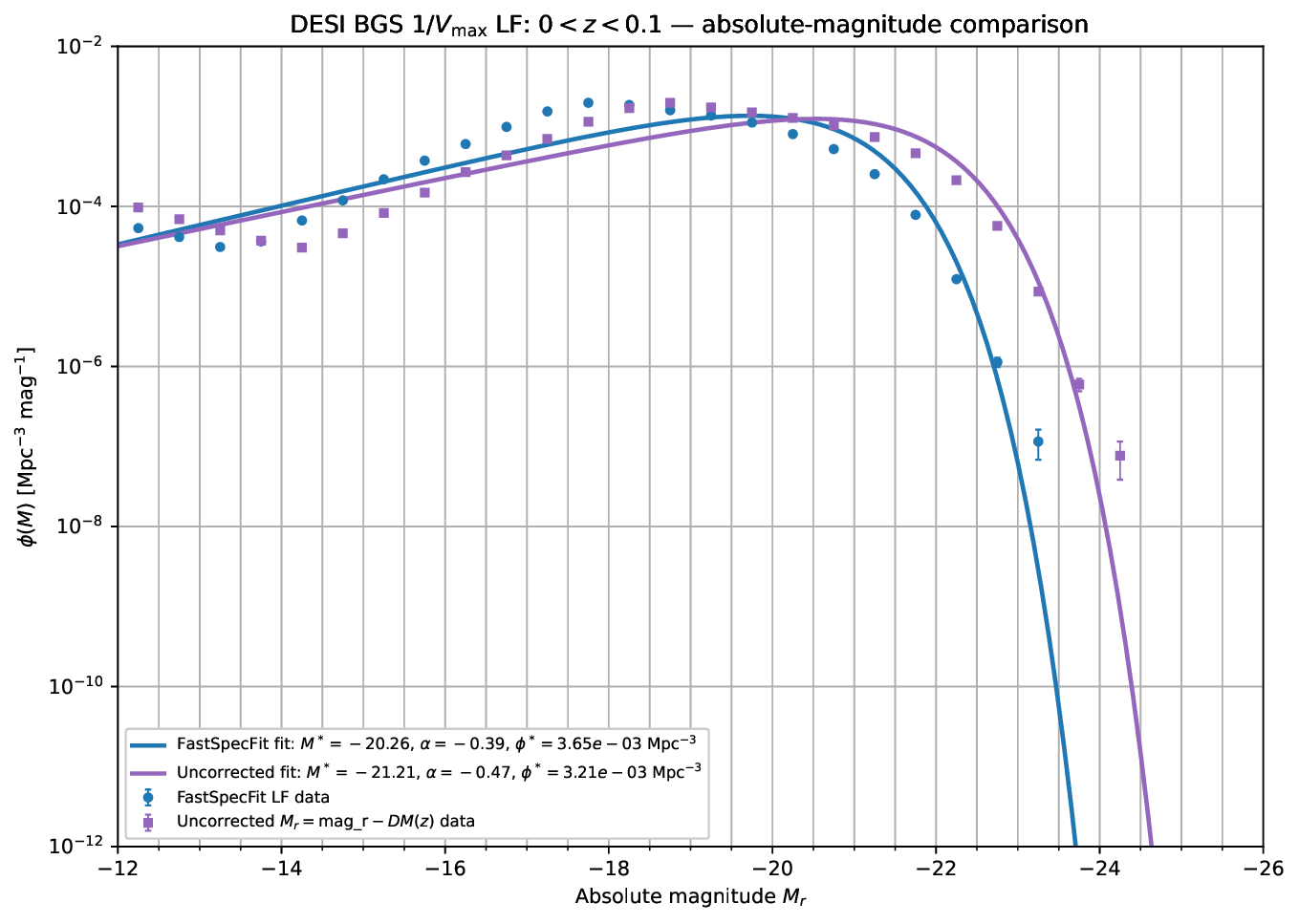}
    &
    \includegraphics[
        width=0.49\textwidth,
        height=0.195\textheight,
        keepaspectratio
    ]{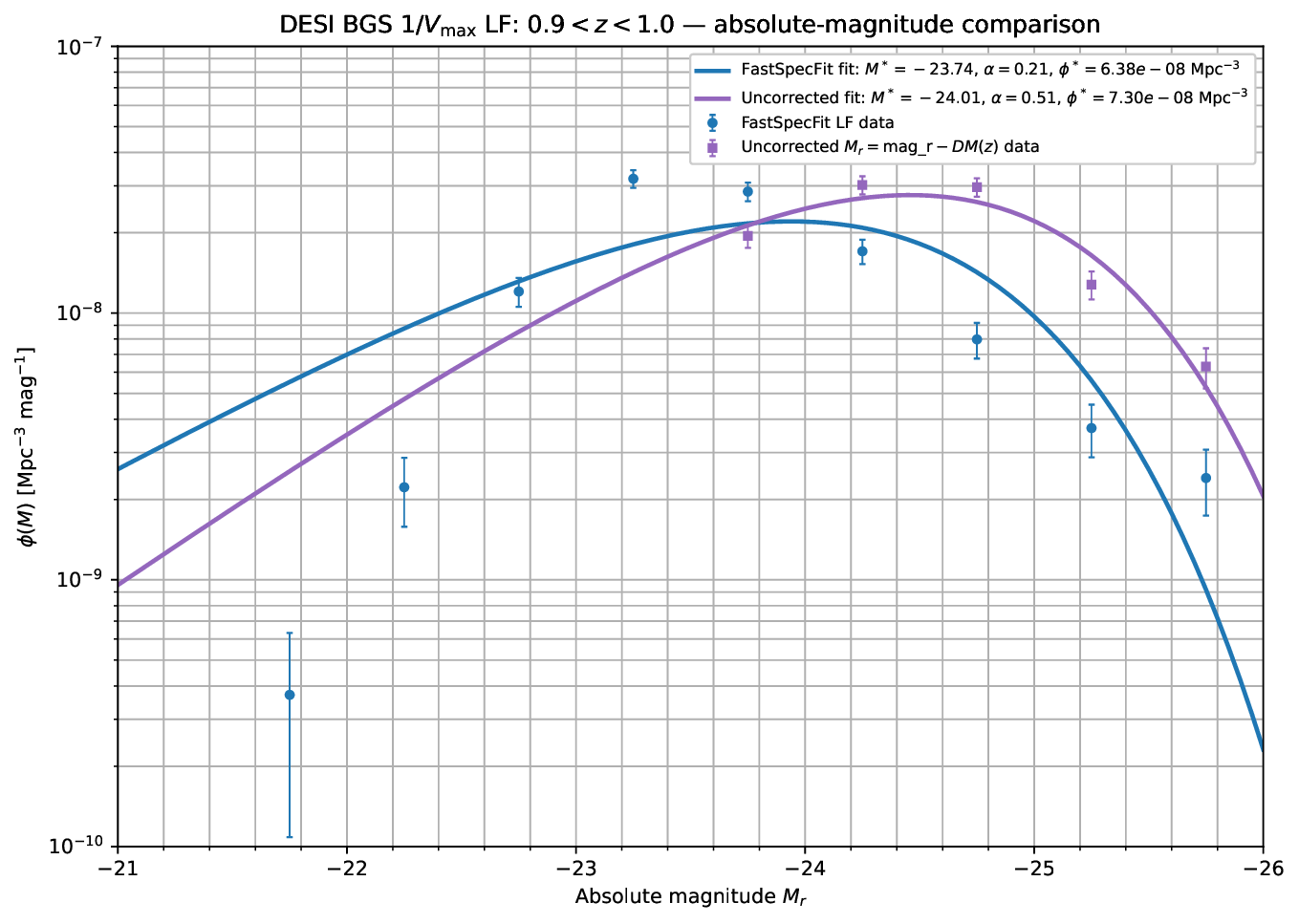}
    \\[-3mm]
    (a) & (b) \\[2mm]

    \includegraphics[
        width=0.49\textwidth,
        height=0.195\textheight,
        keepaspectratio
    ]{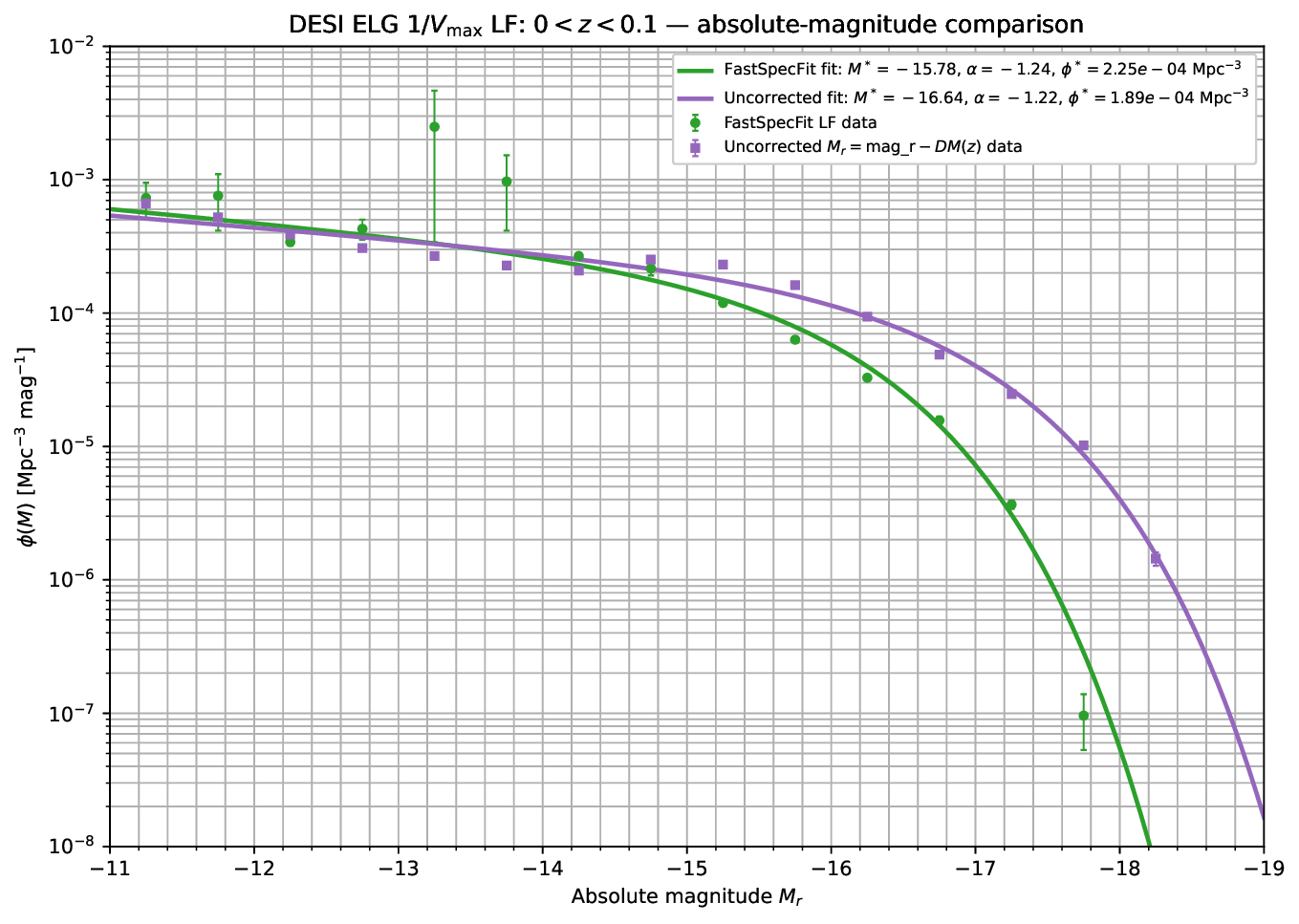}
    &
    \includegraphics[
        width=0.49\textwidth,
        height=0.195\textheight,
        keepaspectratio
    ]{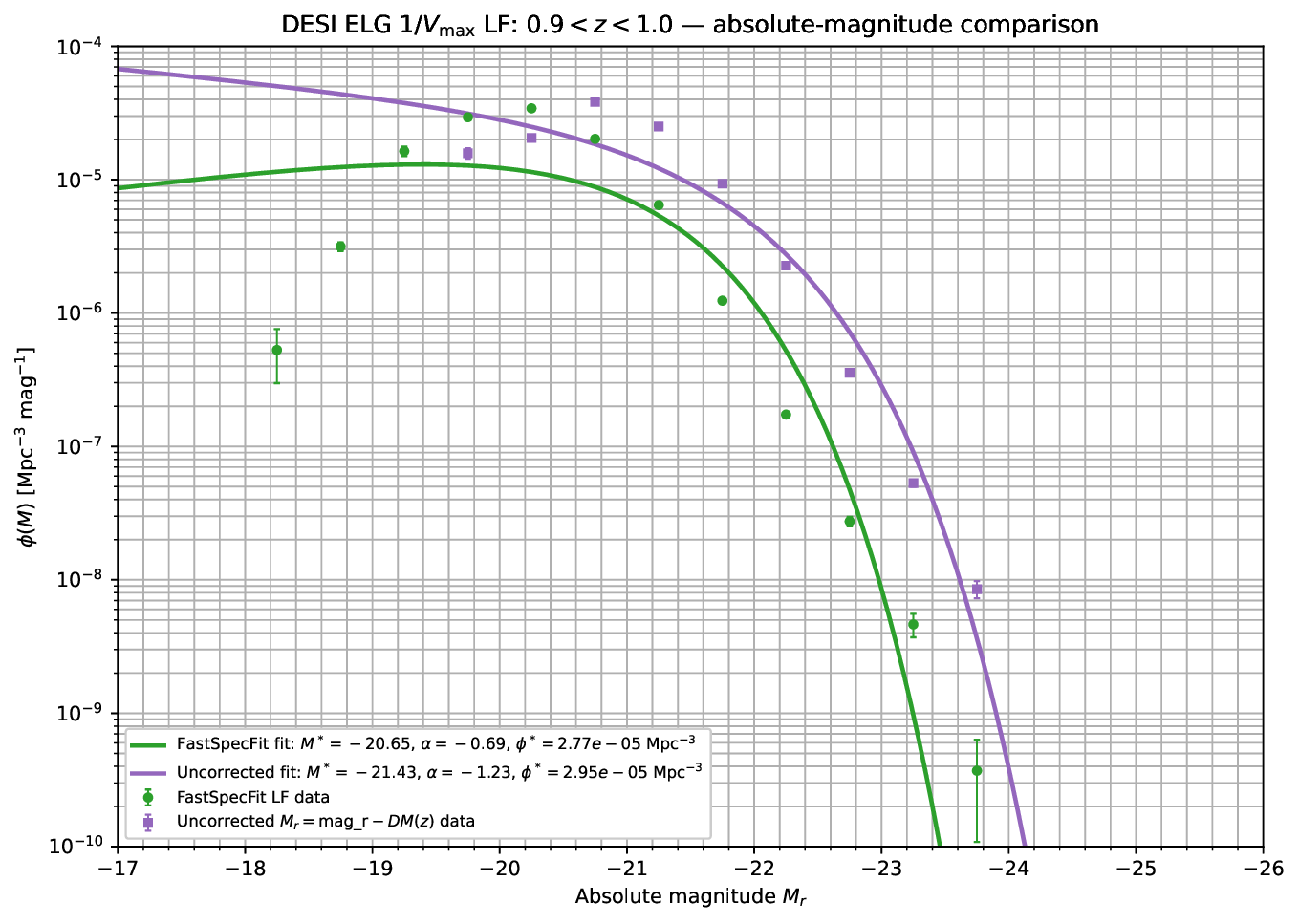}
    \\[-3mm]
    (c) & (d) \\[2mm]

    \includegraphics[
        width=0.49\textwidth,
        height=0.195\textheight,
        keepaspectratio
    ]{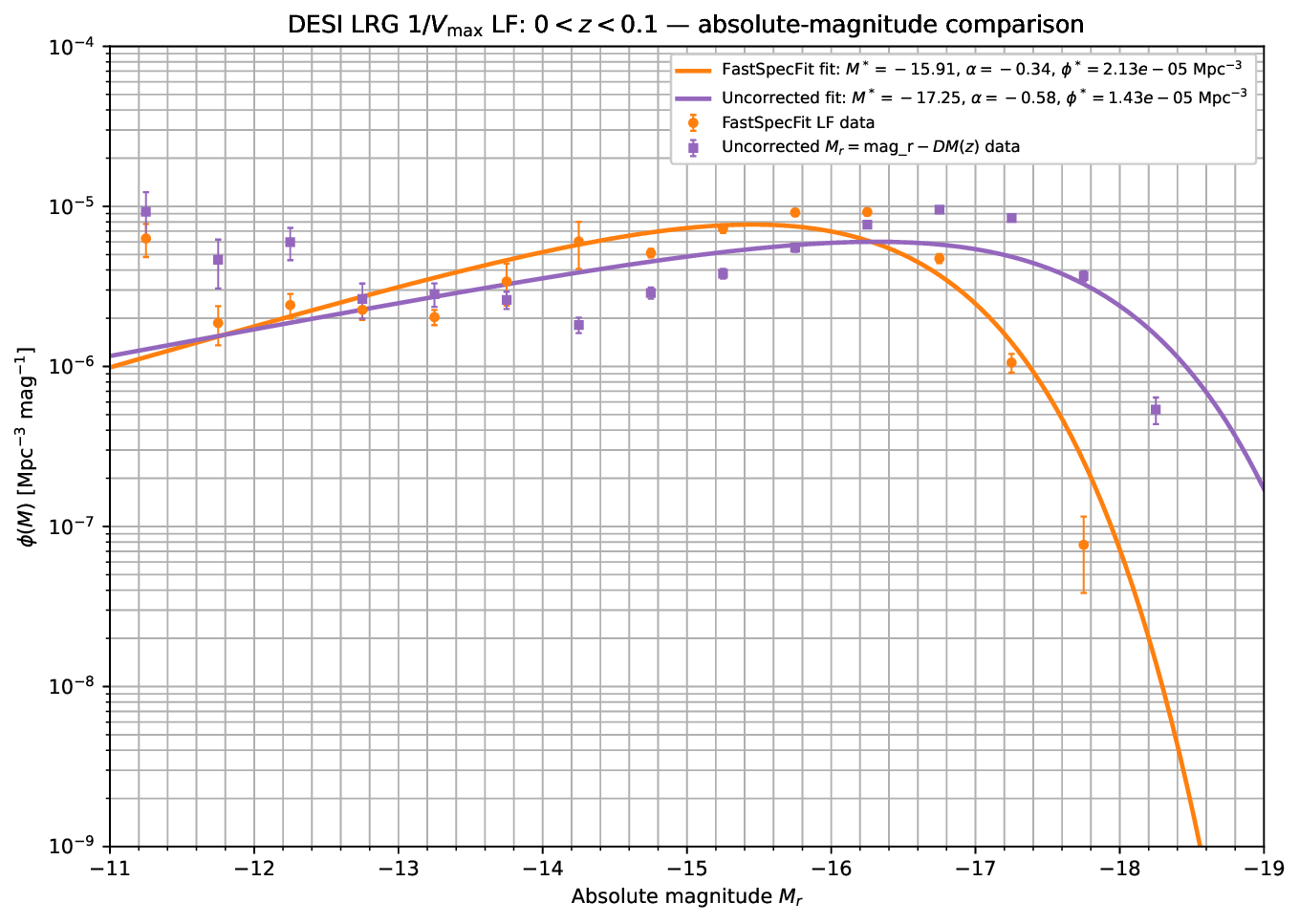}
    &
    \includegraphics[
        width=0.49\textwidth,
        height=0.195\textheight,
        keepaspectratio
    ]{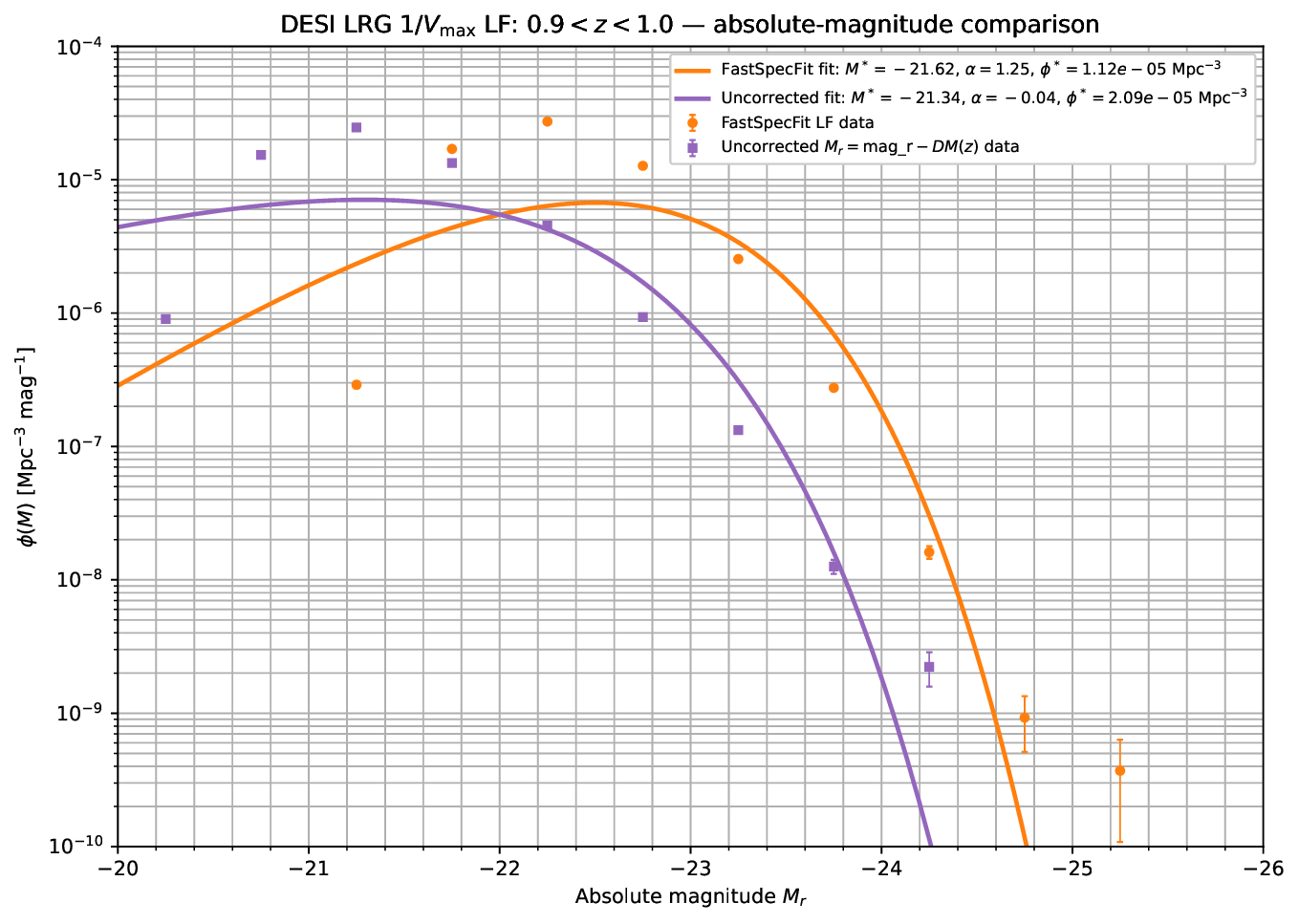}
    \\[-3mm]
    (e) & (f) \\[2mm]

    \includegraphics[
        width=0.49\textwidth,
        height=0.195\textheight,
        keepaspectratio
    ]{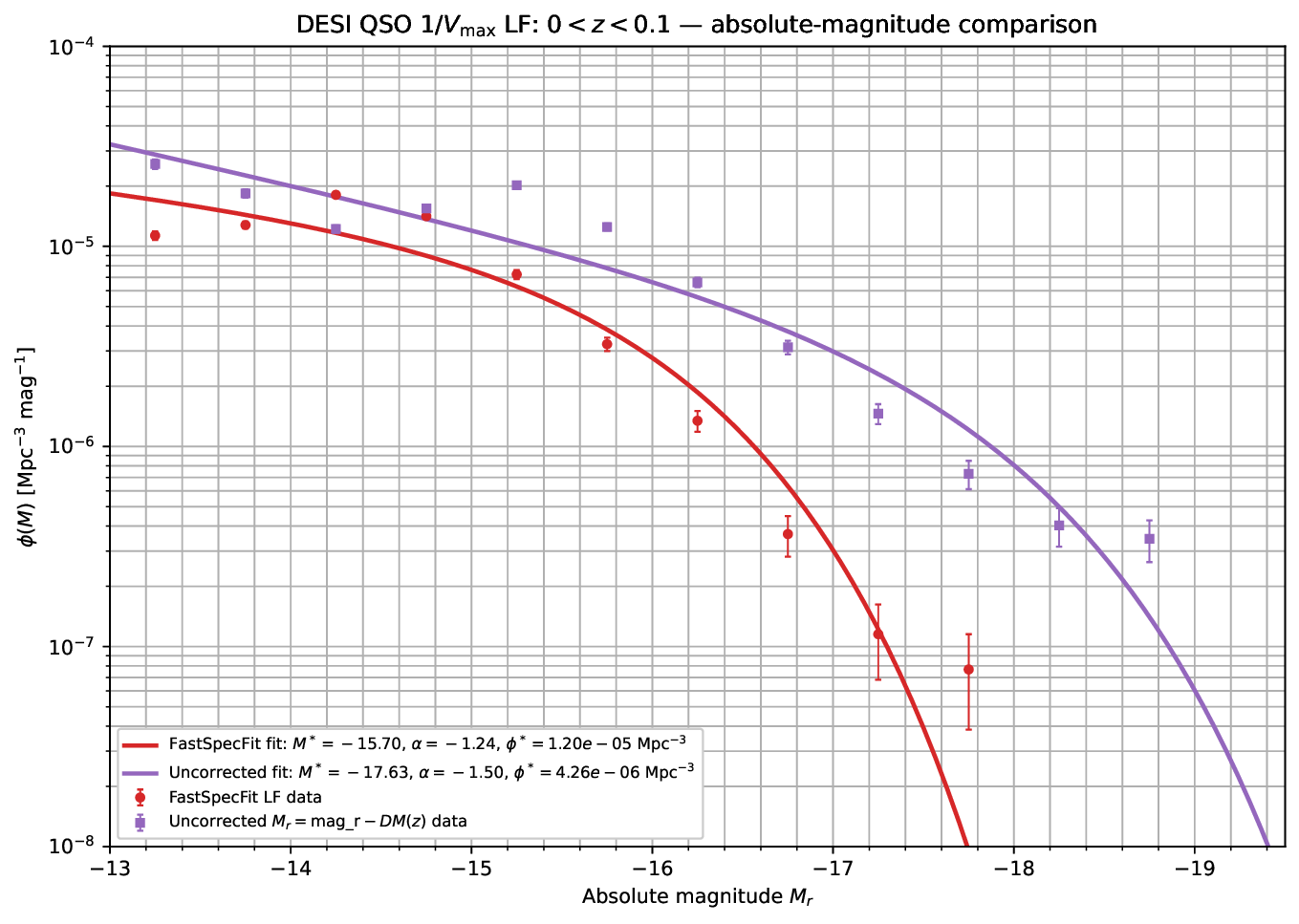}
    &
    \includegraphics[
        width=0.49\textwidth,
        height=0.195\textheight,
        keepaspectratio
    ]{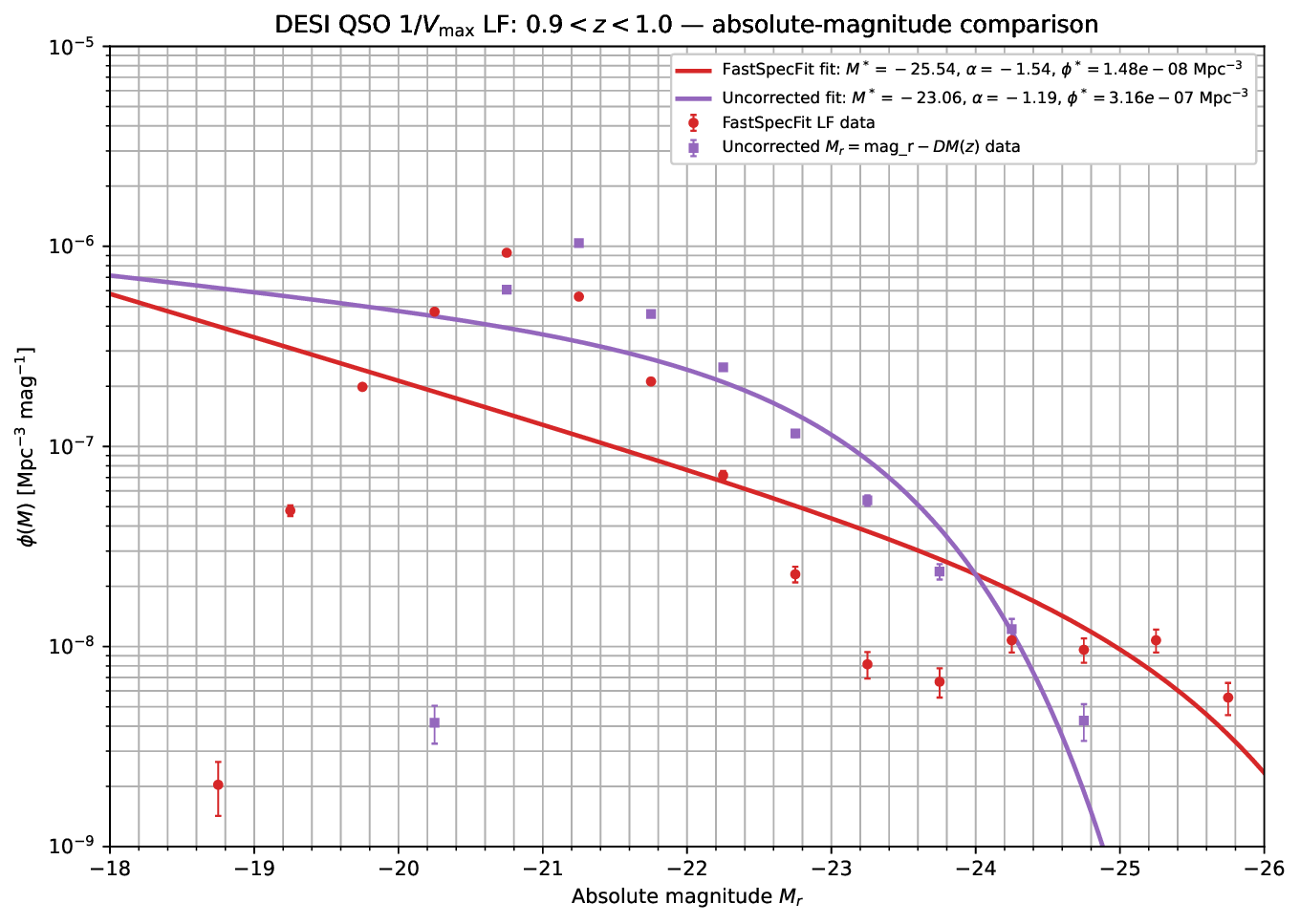}
    \\[-3mm]
    (g) & (h)

    \end{tabular}

\caption{Comparison between FastSpecFit-based corrected and uncorrected $1/V_{\mathrm{max}}$ luminosity functions across DESI target classes and redshift regimes. From top to bottom, rows correspond to the BGS (panels a, b), ELG (panels c, d), LRG (panels e, f), and QSO (panels g, h) target samples. The left column shows the low-redshift baseline ($0.0 < z < 0.1$), while the right column corresponds to intermediate redshifts ($0.9 < z < 1.0$).}



    \label{fig:LF_corrections_targetclass_mosaic}
\end{figure*}

\subsection{Redshift-resolved Luminosity Functions}

Before examining the redshift evolution of individual sample fits, it is instructive to evaluate the direct impact of the baseline corrections on the measured space densities. Figure~\ref{fig:LF_corrections_targetclass_mosaic} presents a systematic comparison between the raw $1/V_{\mathrm{max}}$ measurements and the FastSpecFit-based corrected luminosity functions across all four DESI target classes (BGS, ELG, LRG, and QSO) at low ($0.0 < z < 0.1$) and intermediate ($0.9 < z < 1.0$) redshifts. 

\begin{figure}[ht]

    \centering

    \gridline{
        \fig{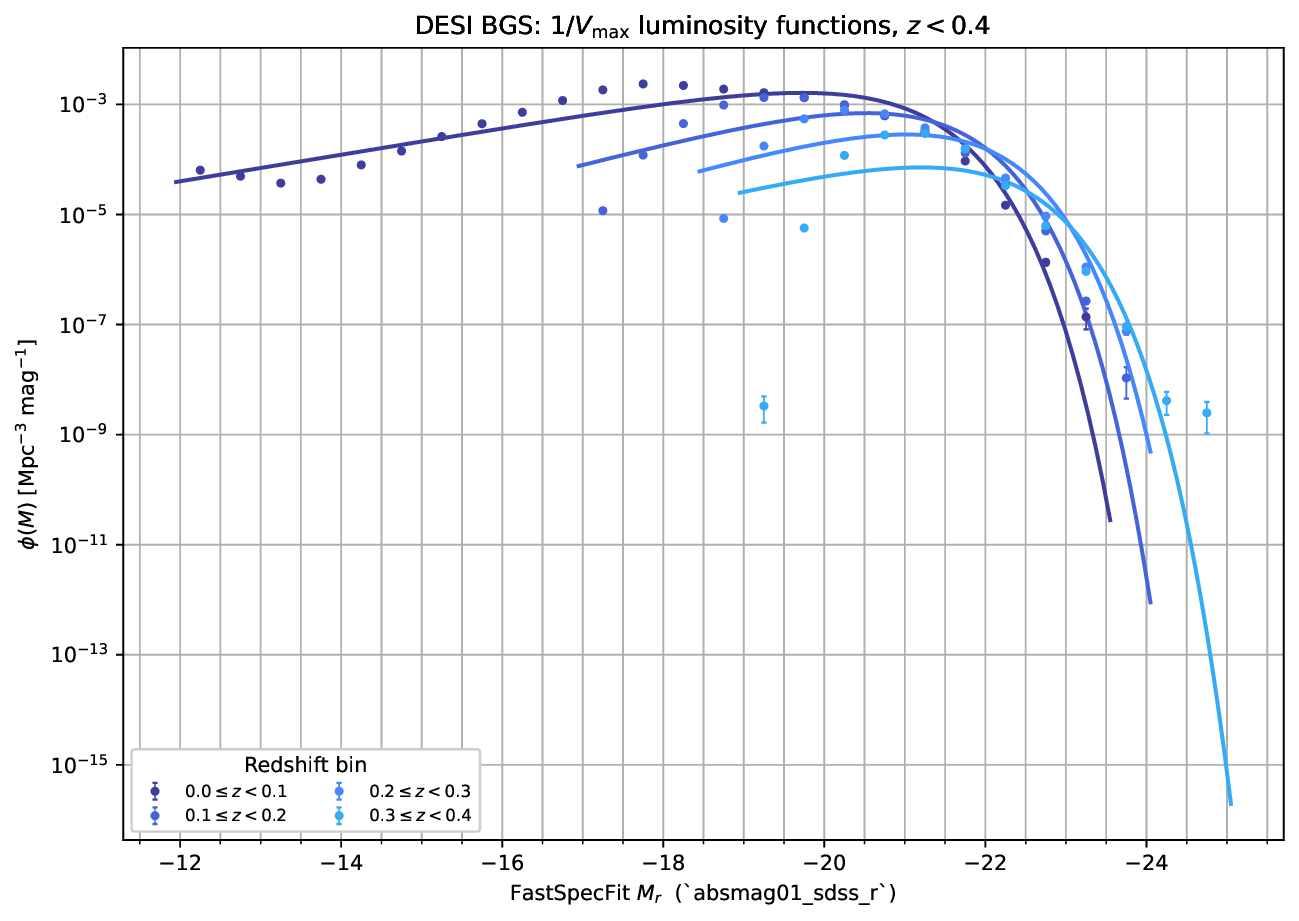}
            {0.32\textwidth}{(a)}
        \fig{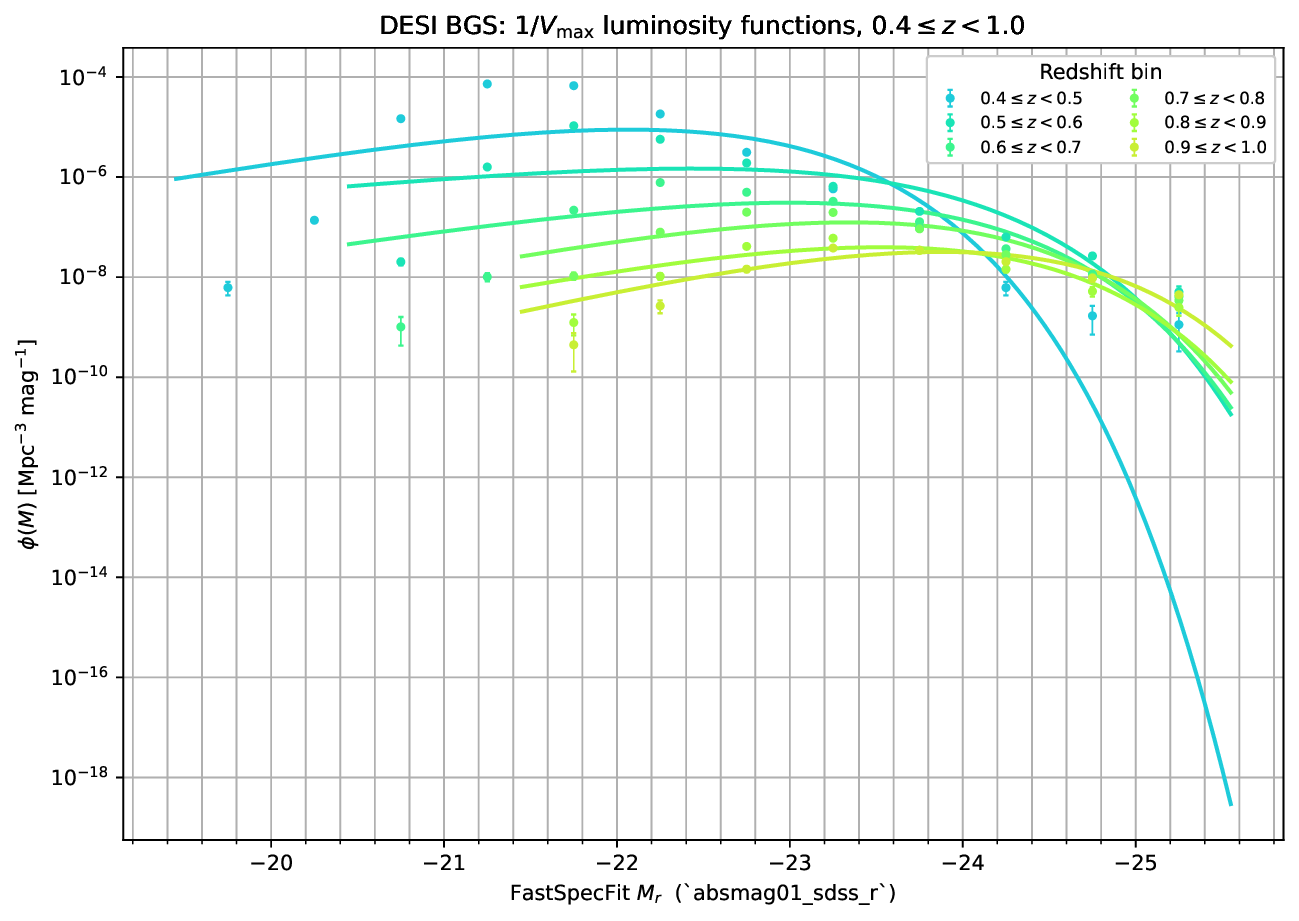}
            {0.32\textwidth}{(b)}
        \fig{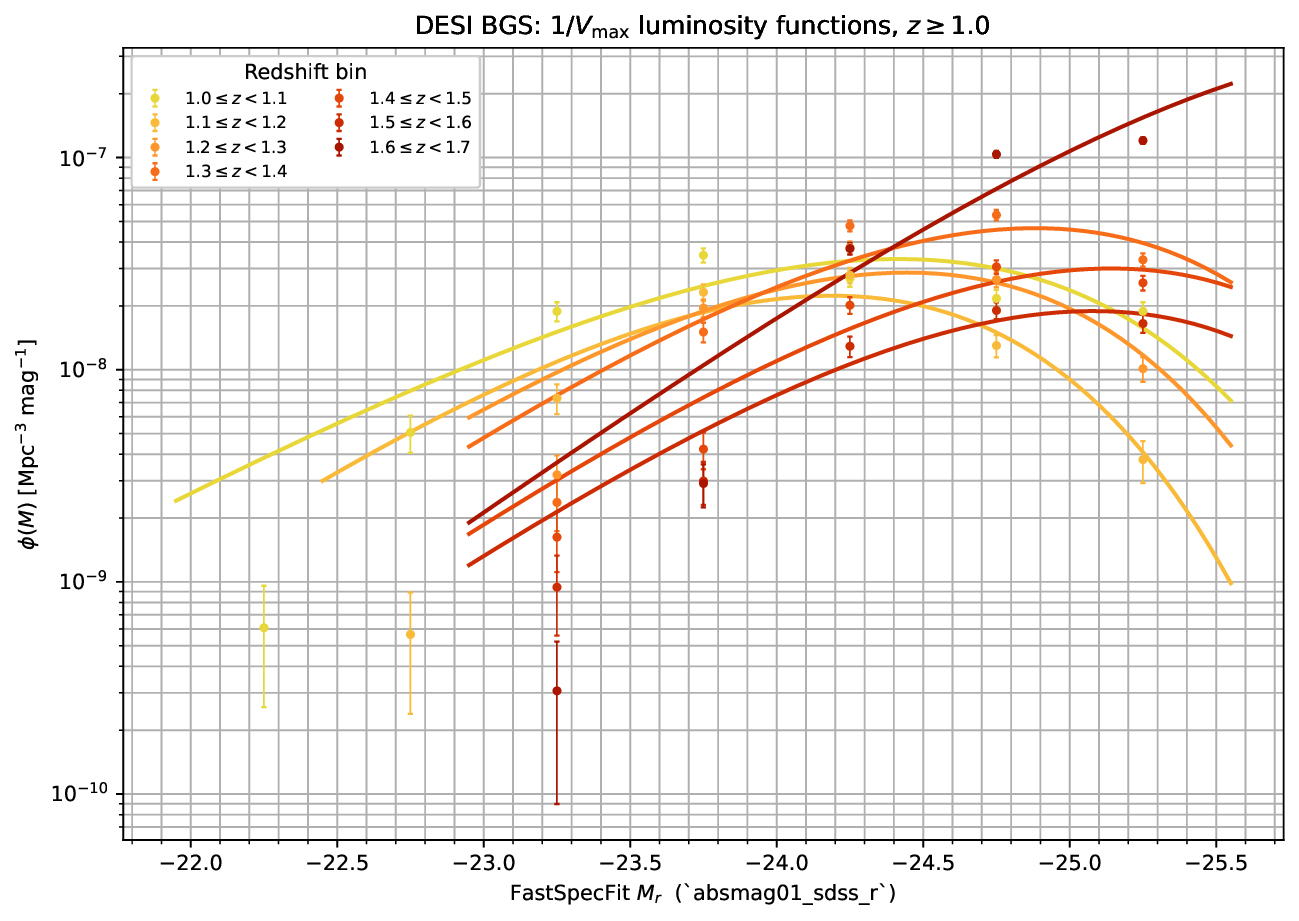}
            {0.32\textwidth}{(c)}
    }

    \caption{Differential galaxy luminosity function $\phi(M)$ for the Bright Galaxy Survey (BGS) target class alone, in redshift bins of width $\Delta z = 0.1$, reconstructed using the $1/V_{\mathrm{max}}$ estimator in three redshift regimes: (a) $z < 0.4$, (b) $0.4 \leq z < 1.0$, and (c) $z \geq 1.0$. The BGS sample dominates at $z<0.4$, exhibiting a steep faint-end slope characteristic of a star-forming-dominated population. The sharp truncation at faint magnitudes in higher-redshift bins marks the effective volume limit of the BGS selection. Error bars represent Poisson uncertainties.}



   \label{fig:lf_bgs}
\end{figure}

\begin{figure}[ht]

    \centering

    \gridline{
        \fig{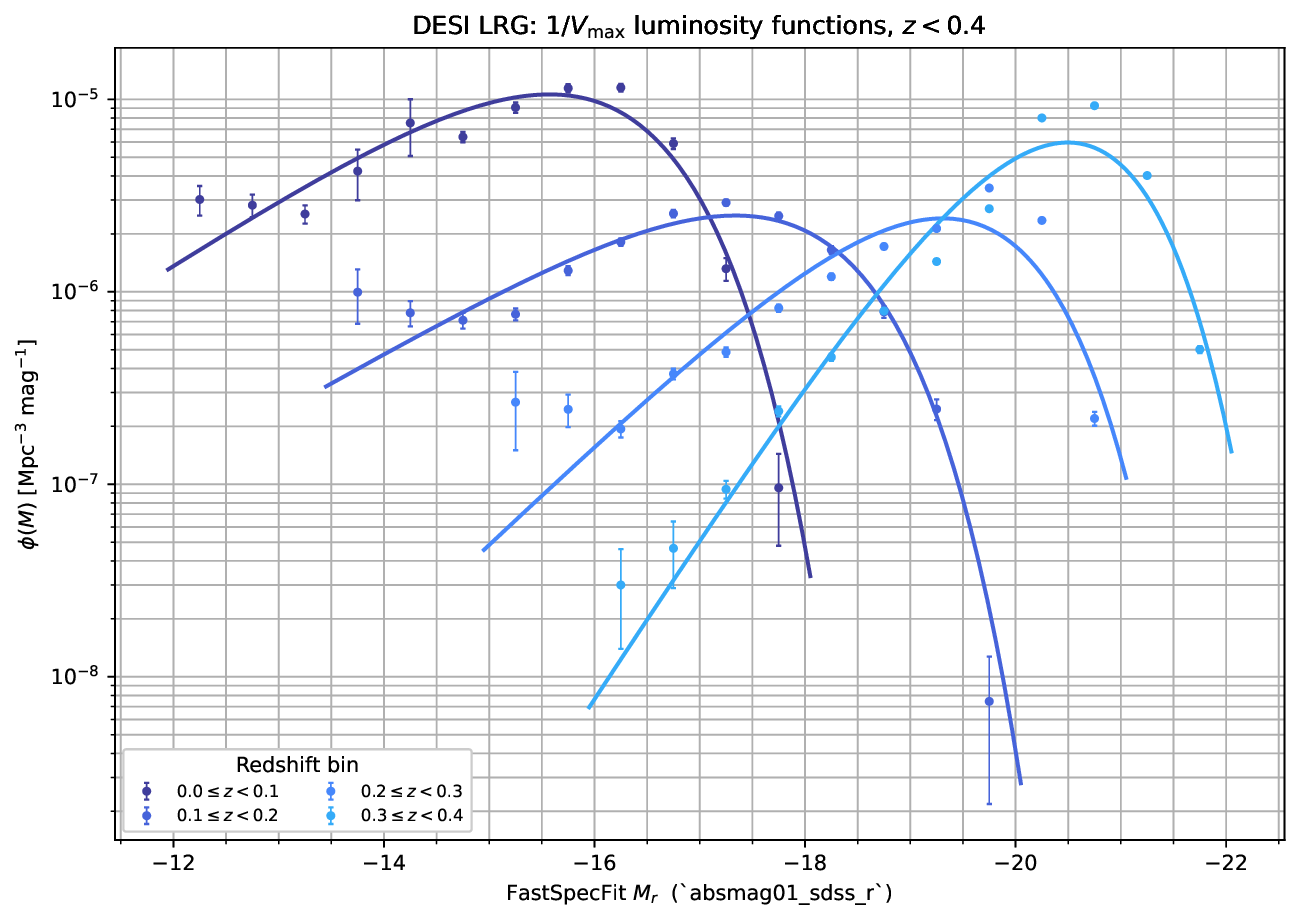}
            {0.32\textwidth}{(a)}
        \fig{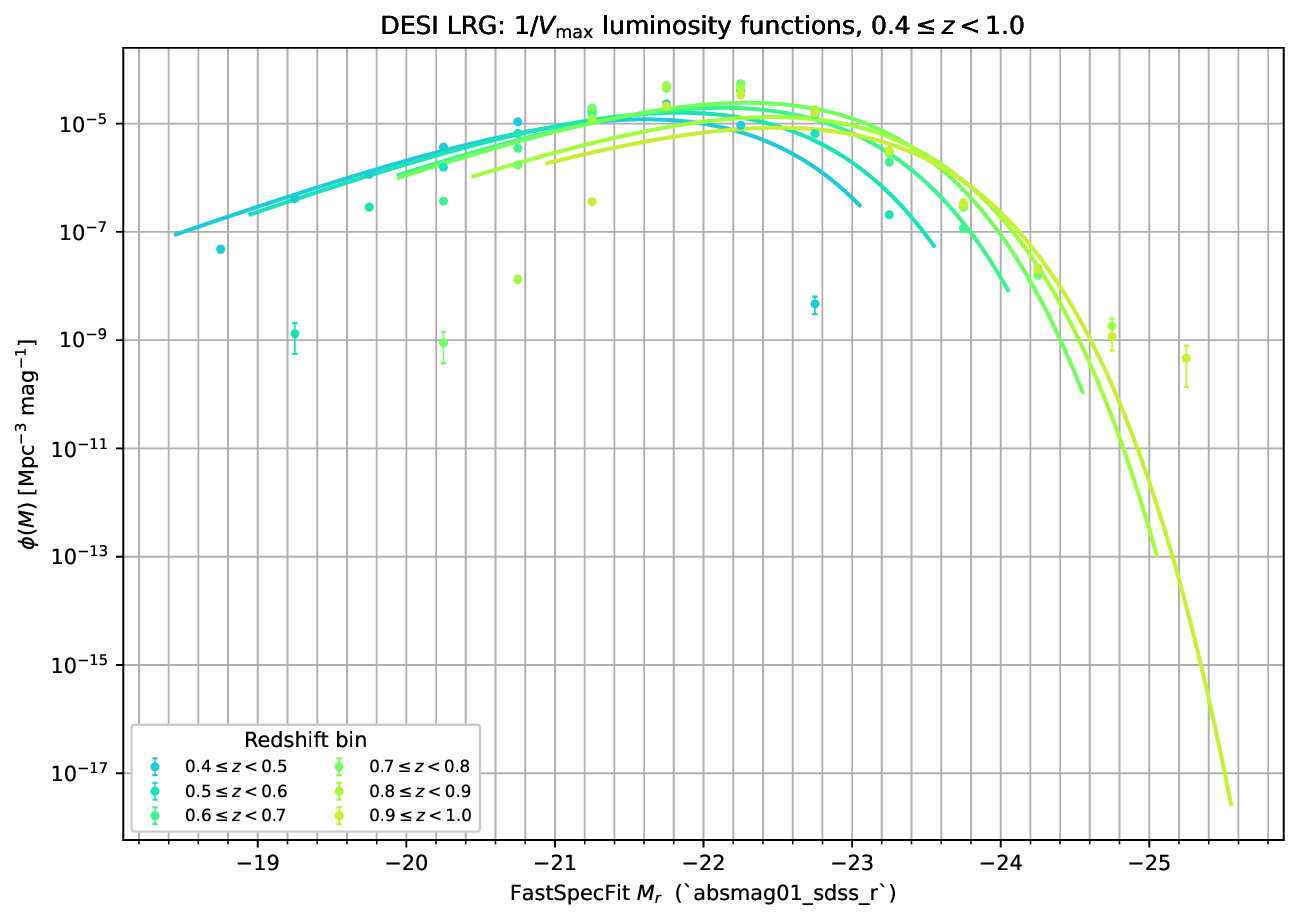}
            {0.32\textwidth}{(b)}
        \fig{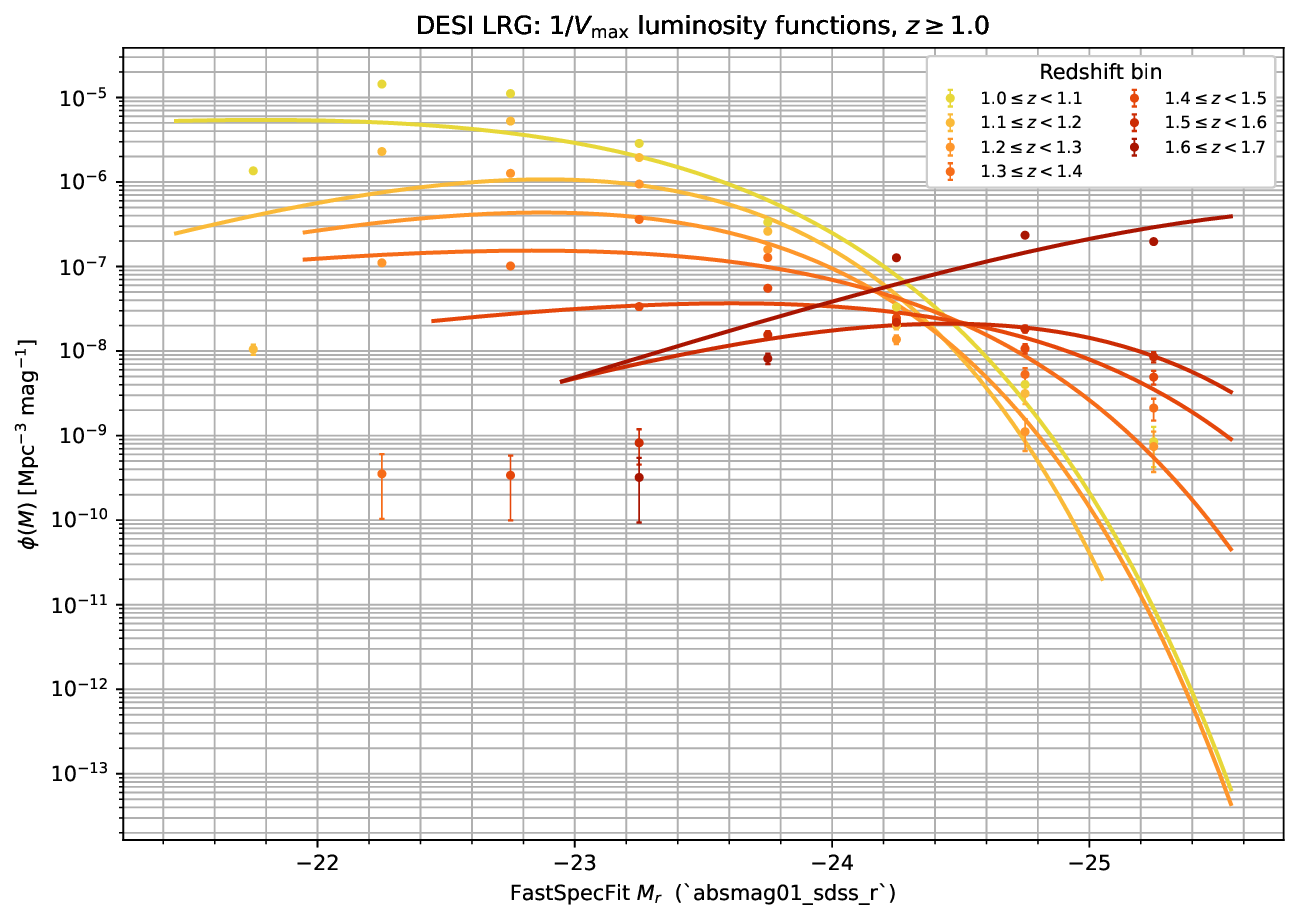}
            {0.32\textwidth}{(c)}
    }

\caption{Differential galaxy luminosity function $\phi(M)$ for the Luminous Red Galaxy (LRG) target class alone, in redshift bins of width $\Delta z = 0.1$, reconstructed using the $1/V_{\mathrm{max}}$ estimator in three redshift regimes:  (a) $z < 0.4$, (b) $0.4 \leq z < 1.0$, and (c) $z \geq 1.0$.. The LRG sample dominates the bright end of the global LF across $0.4<z<1.0$, displaying a truncated luminosity distribution that produces an anomalous, inverted faint-end slope ($\alpha \approx +1.5$) and an artificially stabilized characteristic magnitude ($M^{*} \approx -21$). Error bars represent Poisson uncertainties.}



    
    \label{fig:lf_lrg}
\end{figure}

\begin{figure}[ht]

    \centering

    \gridline{
        \fig{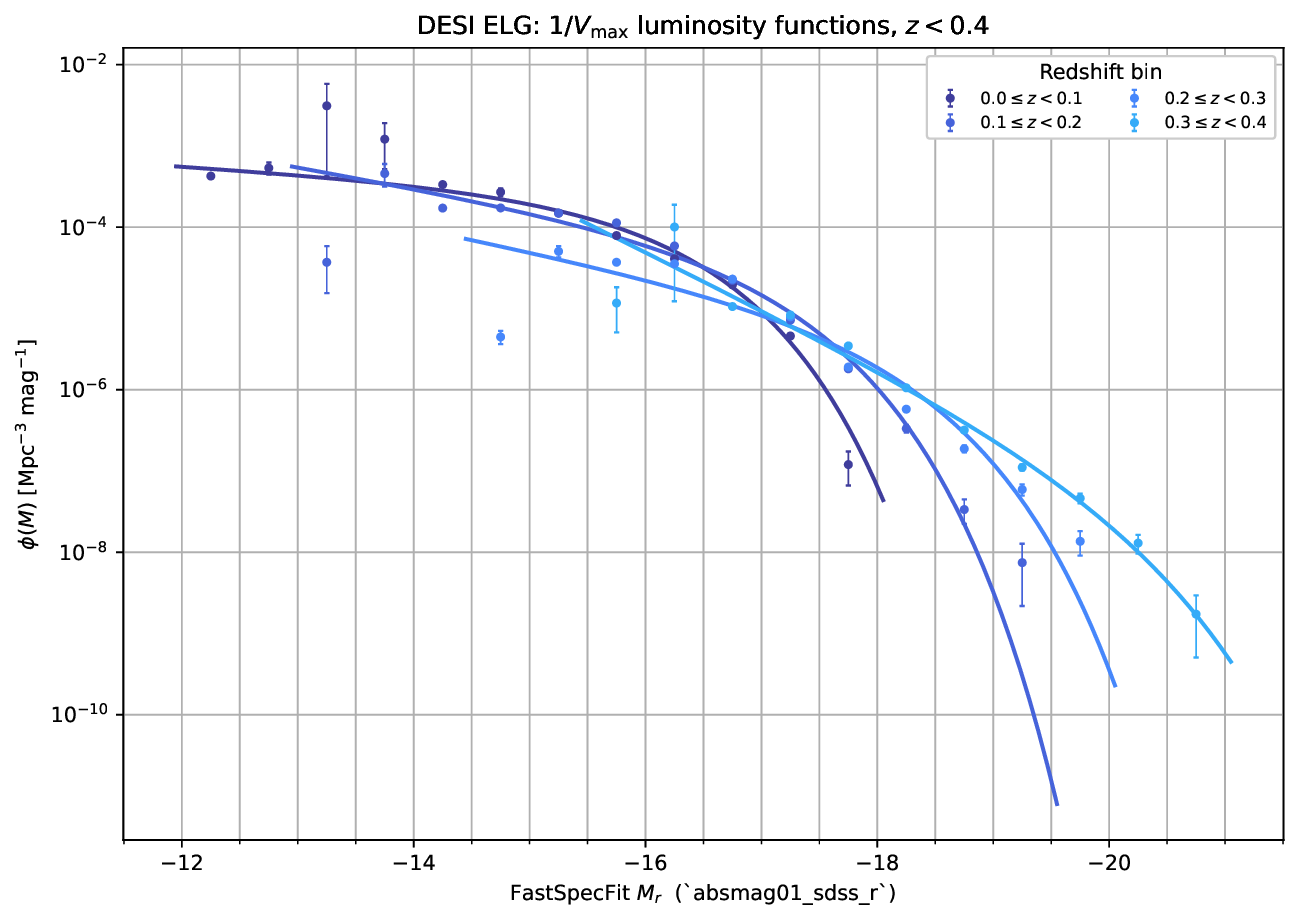}
            {0.32\textwidth}{(a)}
        \fig{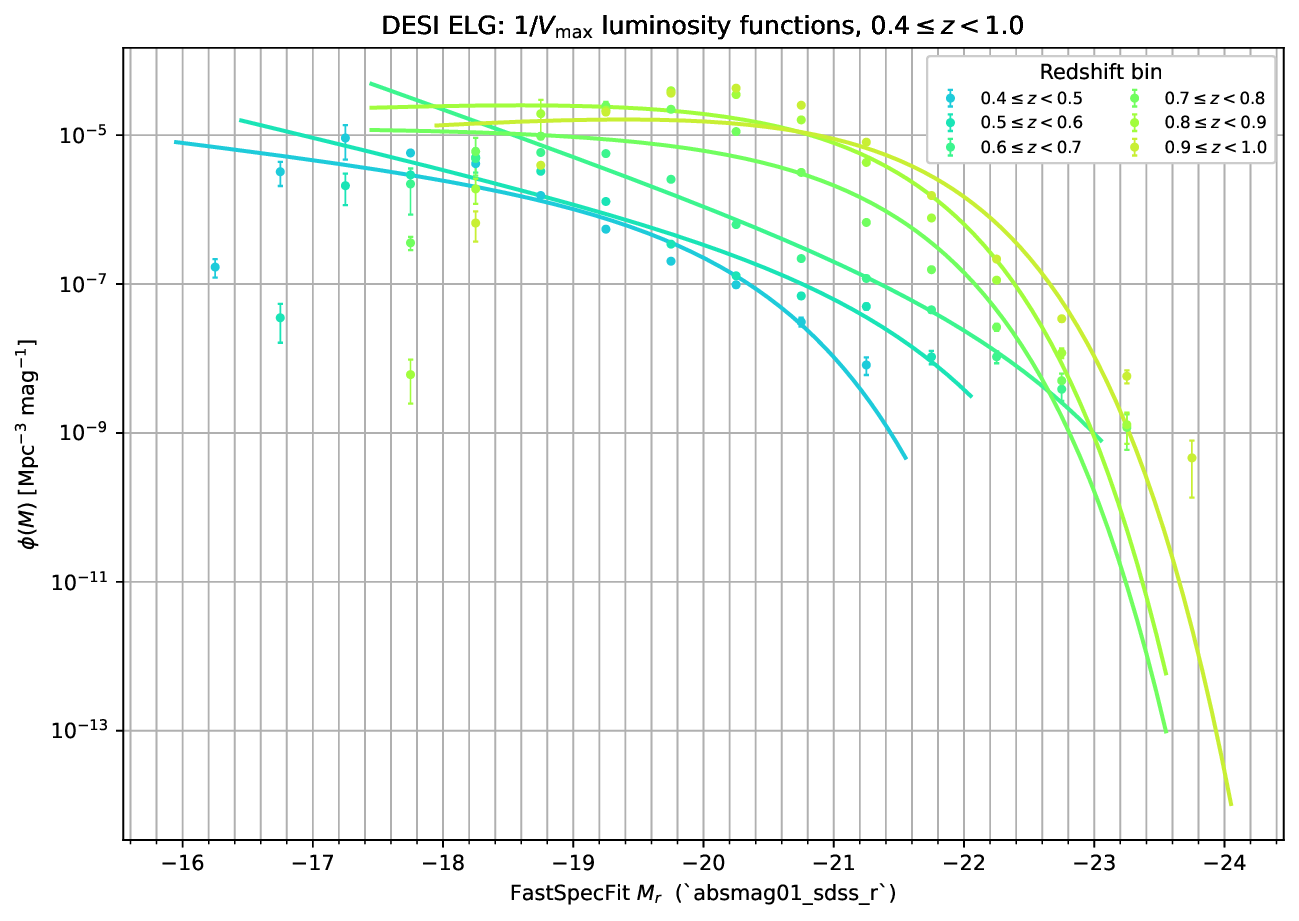}
            {0.32\textwidth}{(b)}
        \fig{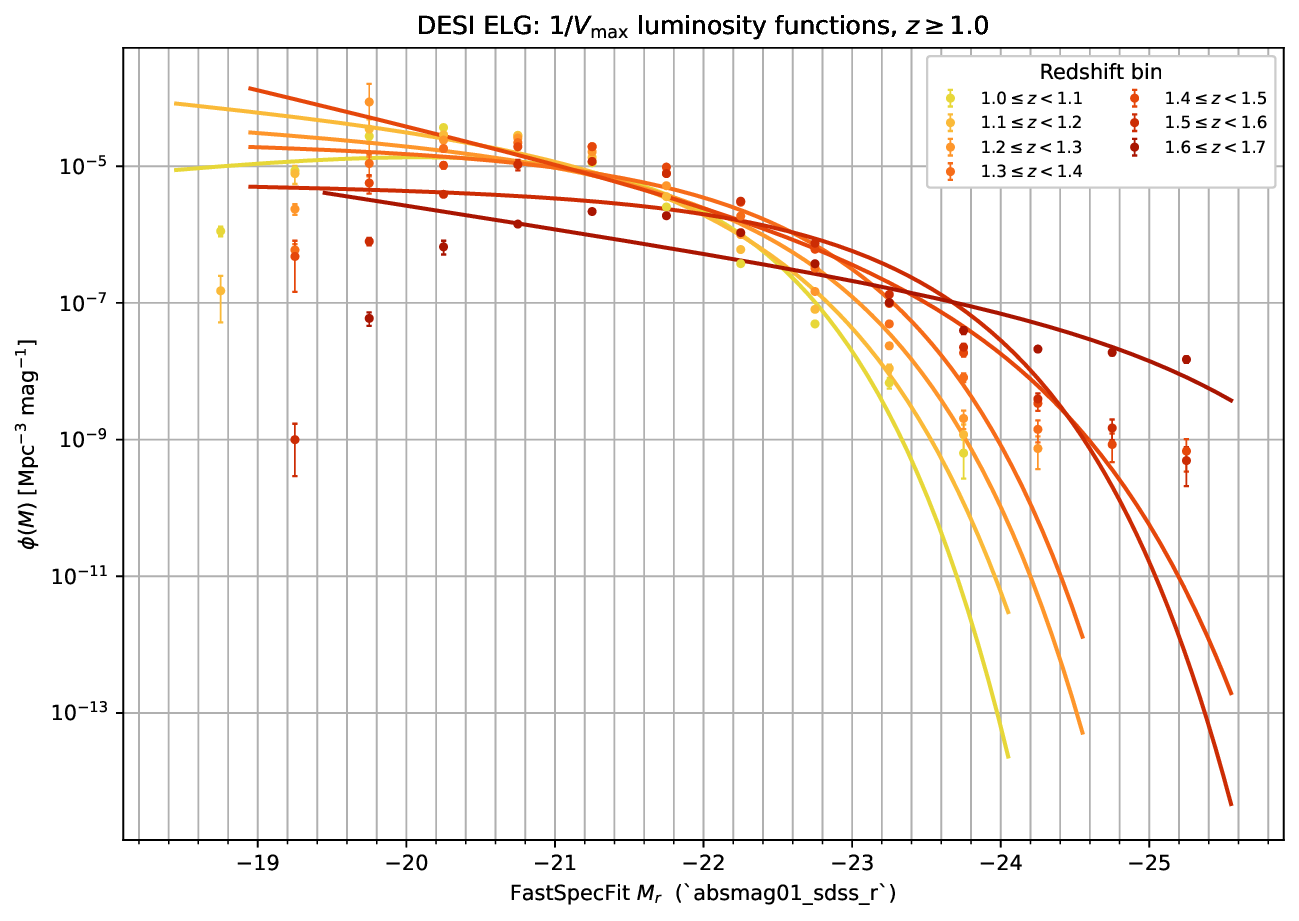}
            {0.32\textwidth}{(c)}
    }

    \caption{Differential galaxy luminosity function \protect{$\phi(M)$} for the Emission Line Galaxy (ELG) target class alone, spanning the range \protect{$0.6 < z < 1.6$}, reconstructed using the \protect{$1/V_{\mathrm{max}}$} estimator in three redshift regimes: (a) $z < 0.4$, (b) $0.4 \leq z < 1.0$, and (c) $z \geq 1.0$. The ELG sample traces actively star-forming galaxies with a steep faint-end slope (\protect{$\alpha \approx -1.0$ to $-1.5$}). At higher redshifts (\protect{$z > 1.0$}), the distributions become increasingly unstable and poorly constrained due to small-number statistics and the progressive loss of extended systems to cosmological \protect{$(1+z)^{-4}$} surface-brightness dimming. Error bars represent Poisson uncertainties.}



    
    \label{fig:lf_elg}
\end{figure}

\begin{figure}[ht]

    \centering

    \gridline{
        \fig{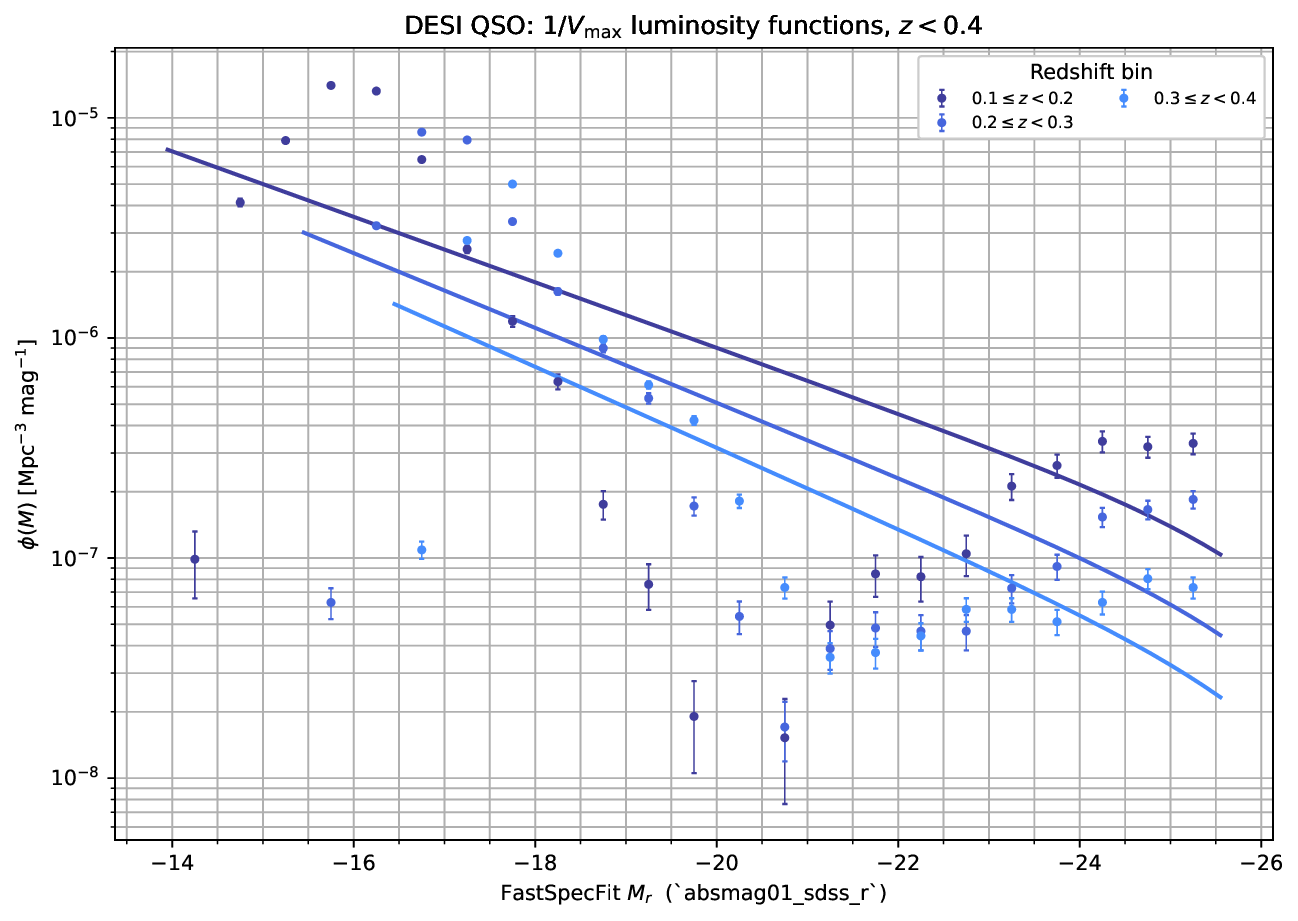}
            {0.32\textwidth}{(a)}
        \fig{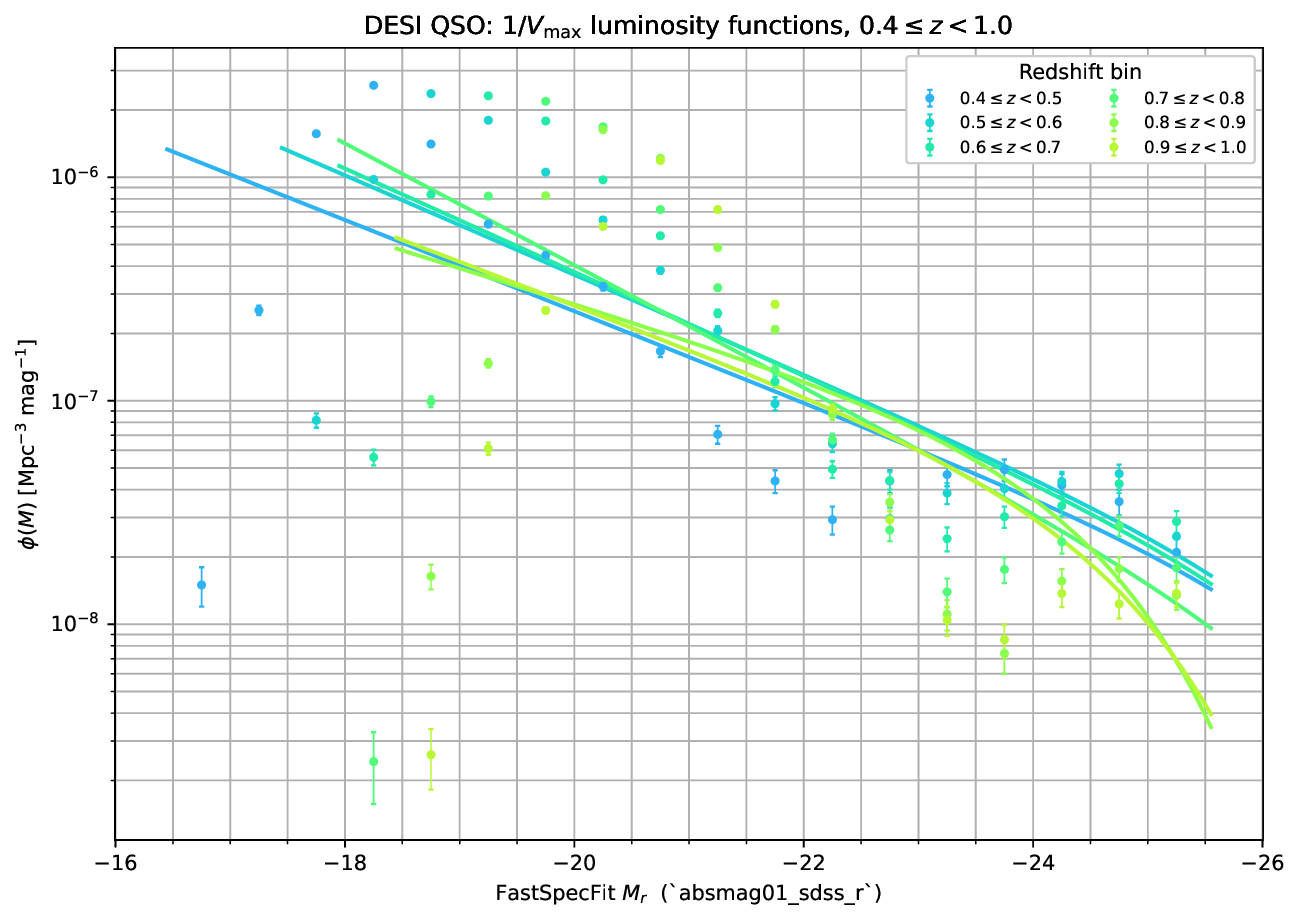}
            {0.32\textwidth}{(b)}
        \fig{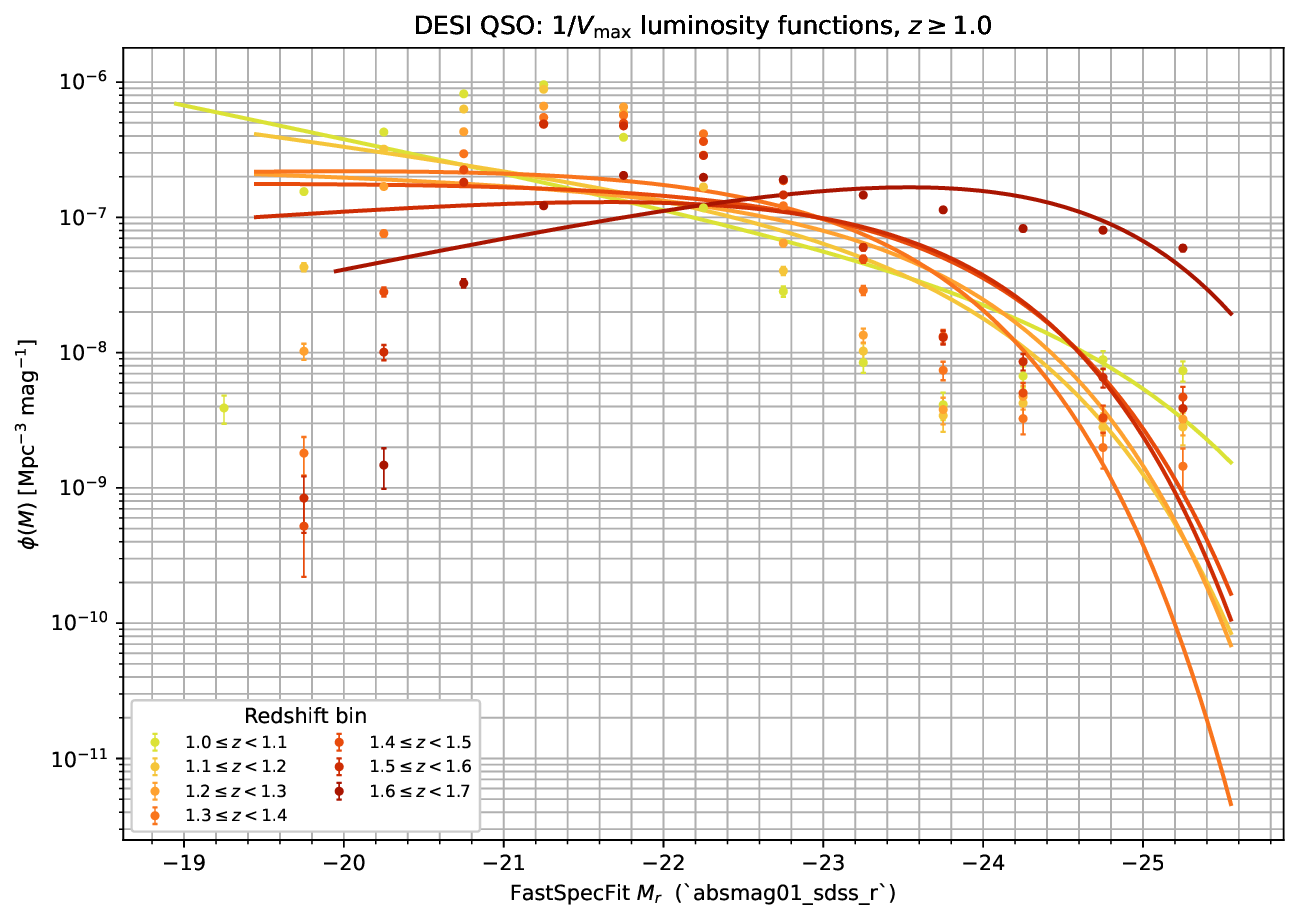}
            {0.32\textwidth}{(c)}
    }

    \caption{Differential galaxy luminosity function $\phi(M)$ for the Quasar (QSO) target class alone, in redshift bins of width $\Delta z = 0.1$, reconstructed using the $1/V_{\mathrm{max}}$ estimator in three redshift regimes: (a) $z < 0.4$, (b) $0.4 \leq z < 1.0$, and (c) $z \geq 1.0$. QSOs occupy extremely bright absolute magnitudes ($M^{*} \protect\sim -24$ to $-28$) and exhibit very low normalization densities ($\phi^{*} \protect\sim 10^{-6}$ to $10^{-8}$). Although not representative of normal galaxy populations, QSOs are included here to illustrate their role in the highest-redshift tail of the DESI DR1 sample. Error bars represent Poisson uncertainties.}



    \label{fig:lf_qso}
\end{figure}

As demonstrated in the panel sequence, applying the three core observational pipeline corrections—Galactic extinction ($A_{\mathrm{MW}}$), rest-frame $K$-corrections, and fiber-to-total aperture normalization—systematically restores the faint-end slopes ($\alpha$) and recovers the density normalization ($\phi^*$) across all target types. The offset between raw and corrected data is particularly severe at intermediate redshifts ($z \sim 0.9\text{--}1.0$) for BGS and ELG samples, where aperture losses and extinction heavily suppress uncorrected flux measurements. Conversely, for compact QSO sources, the correction primarily induces a shift along the absolute magnitude axis ($M^*$). In line with our diagnostic framework (Section~\ref{sec:limitations}), these curves represent baseline corrections without external model-dependent adjustments (such as cosmic evolution $e(z)$ or inclination cuts). Having established the quantitative scale of these baseline corrections, the following subsections detail the resulting luminosity functions and diagnostic Schechter parameters derived for each target class across the full redshift range (Figures~\ref{fig:lf_bgs} through \ref{fig:lf_qso}).

\begin{figure}[ht]

    \centering

    \gridline{
        \fig{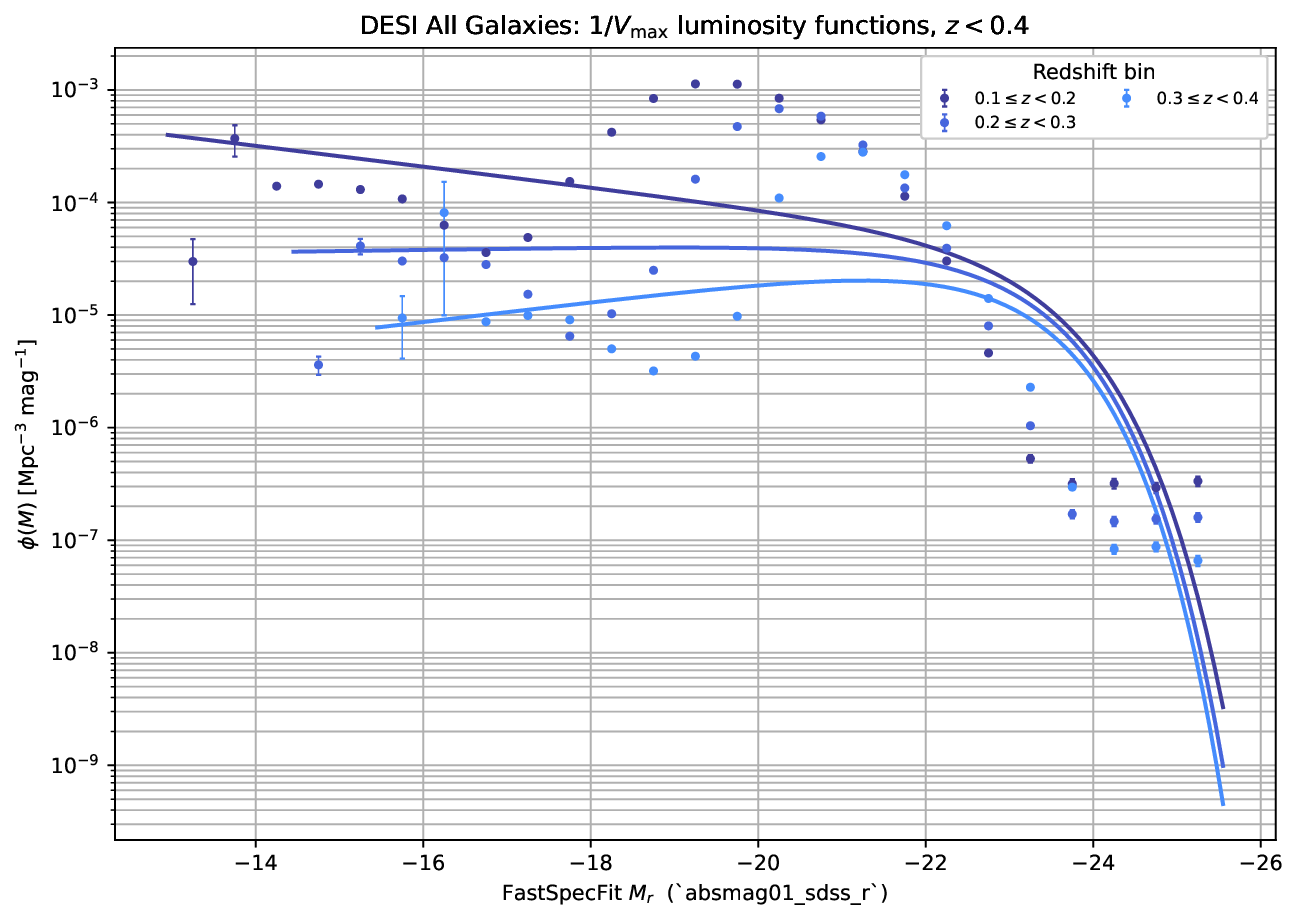}
            {0.32\textwidth}{(a)}
        \fig{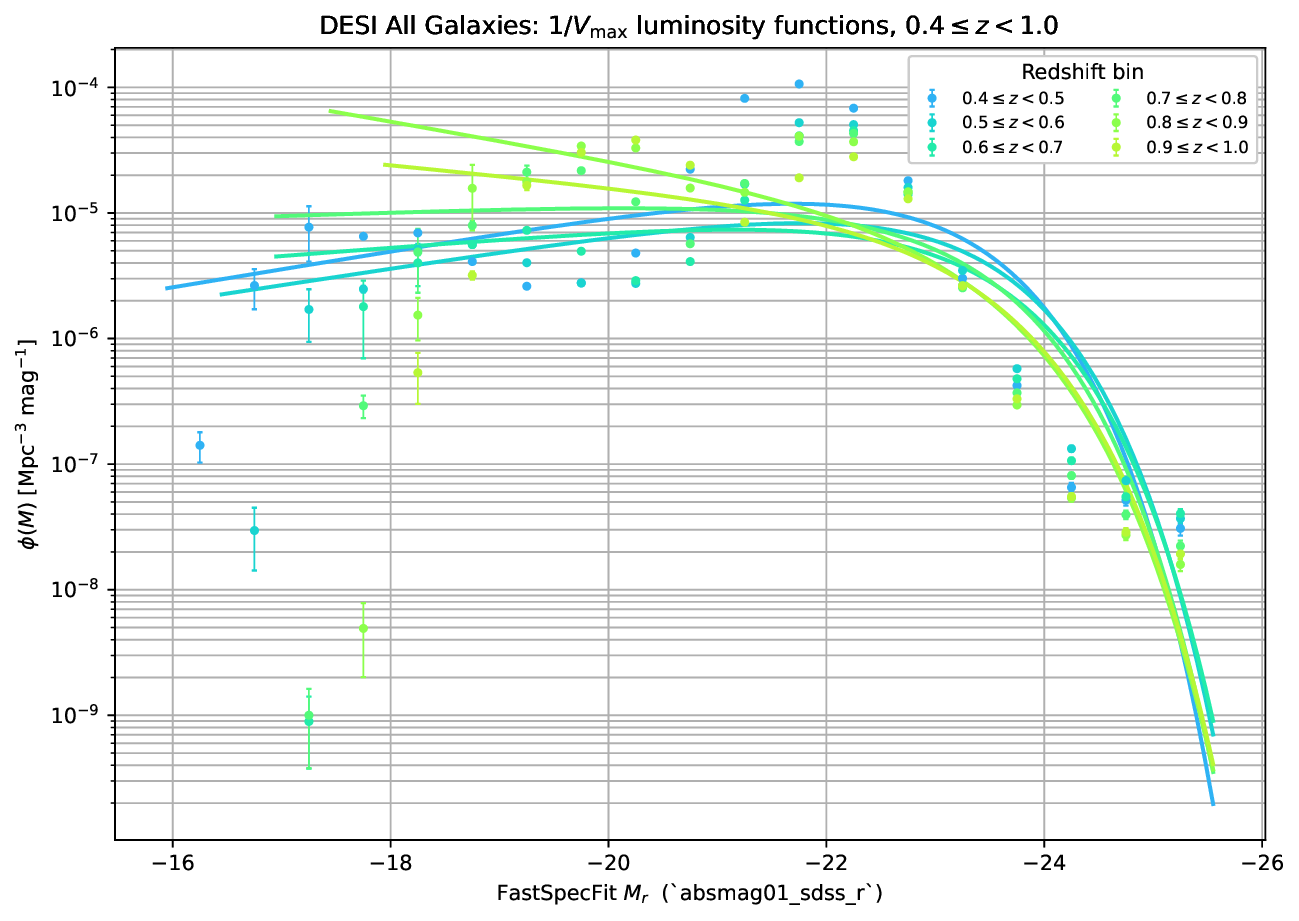}
            {0.32\textwidth}{(b)}
        \fig{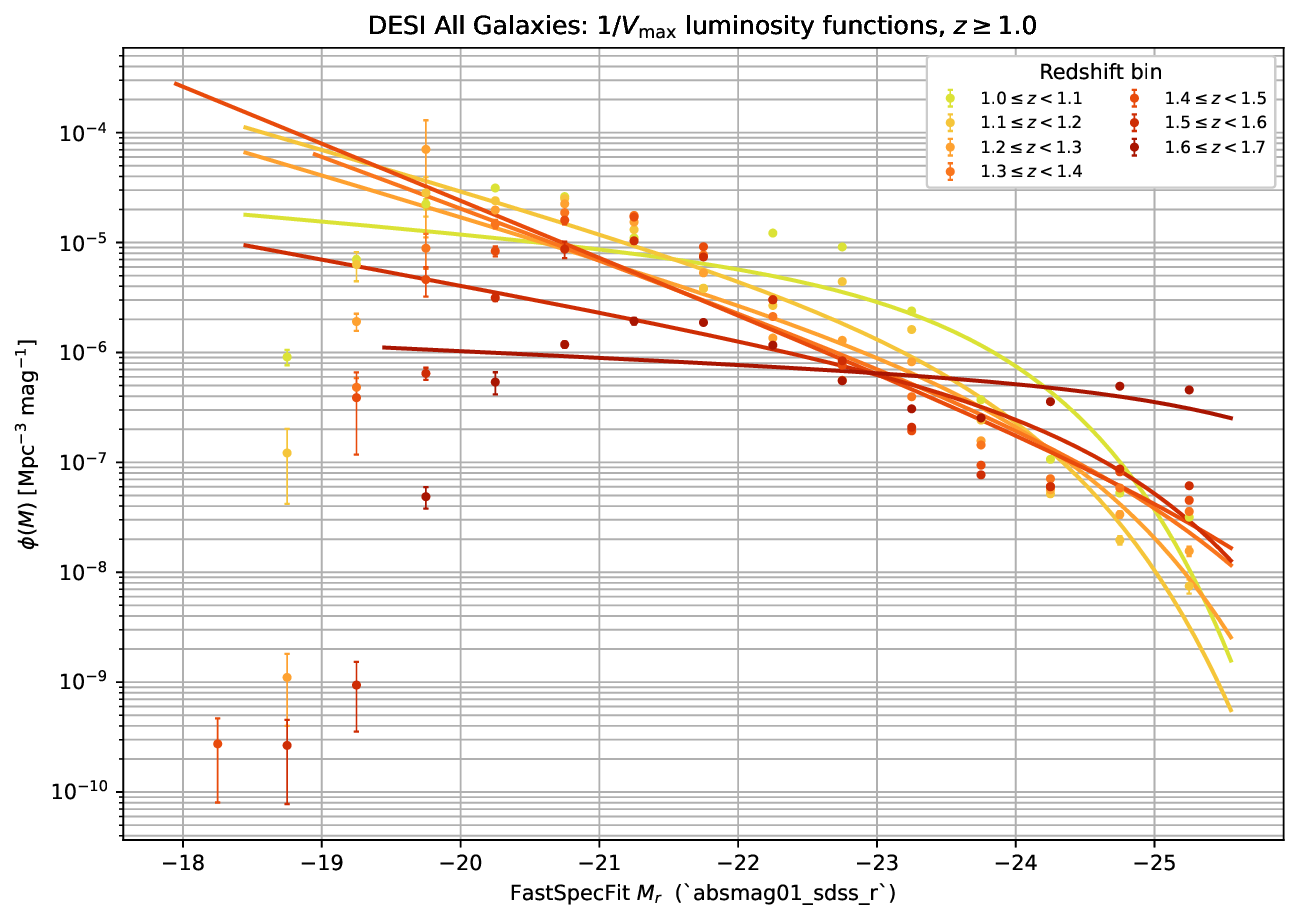}
            {0.32\textwidth}{(c)}
    }

\caption{Differential galaxy luminosity function $\phi(M)$ for the combined DESI DR1 sample in redshift bins of width $\Delta z = 0.1$, reconstructed using the $1/V_{\mathrm{max}}$ estimator in three redshift regimes:  (a) $z < 0.4$, (b) $0.4 \leq z < 1.0$, and (c) $z \geq 1.0$. The global LF is dominated by different target classes at different redshifts: BGS at $z<0.4$, LRG and ELG at $0.4<z<0.8$, and only the brightest ELGs and QSOs beyond $z>1.0$. Error bars represent Poisson uncertainties.}



    \label{fig:lf_combined}
\end{figure}

Figure~\ref{fig:lf_combined} presents the empirical differential galaxy luminosity function, $\phi(M)$, for the combined DESI DR1 sample across independent redshift bins of width $\Delta z = 0.1$, reconstructed using the non-parametric $1/V_{\mathrm{max}}$ estimator and structured across three panels corresponding to distinct redshift regimes ($z < 0.4$, $0.4 \le z < 1.0$, and $z \ge 1.0$). Figures~\ref{fig:lf_bgs},~\ref{fig:lf_lrg},~\ref{fig:lf_elg}, and~\ref{fig:lf_qso} display the corresponding distributions evaluated separately for the BGS, LRG, ELG, and QSO target classes, respectively. This multi-panel representation clearly highlights that the empirical data exhibit severe, high-significance modulations relative to the smooth parametric Schechter fits. Rather than reflecting intrinsic cosmological evolution, these extreme departures mark the imprint of target-selection completeness transitions, fiber assignment boundaries, and uncorrected large-scale structure features in DESI DR1. Furthermore, we deliberately retain and fit the luminosity function bins extending into both the faint- and bright-end incomplete regimes. While faint-end incompleteness is driven by Malmquist bias and target detection thresholds, bright-end incompleteness arises from photometric saturation masks, explicit bright-magnitude selection limits in DESI targets (e.g., upper boundary cuts in BGS), and finite volume sampling of rare, highly luminous systems. Because these parametric fits serve a strictly diagnostic role—mapping the exact thresholds where observational completeness breaks down—retaining these boundary bins explicitly illustrates the operational limits of the survey sample.

At $z < 0.4$ (Figure~\ref{fig:lf_combined}), the morphology of the global LF is heavily dominated by the BGS population, displaying a well-sampled bright-end cutoff and a steep faint end. This structure matches the BGS-only LF shown in Figure~\ref{fig:lf_bgs}, which exhibits a steep faint-end slope before undergoing a sharp, volume-limited truncation at faint magnitudes dictated by the survey's apparent magnitude limits.

Beyond $z \approx 0.4$, the BGS contribution drops abruptly as the population falls below the nominal flux detection thresholds. In the intermediate interval ($0.4 < z < 1.0$), the global LF ceases to represent a single physical population, transforming instead into a mathematical superposition of structurally distinct components. Here, the LRG sample (Figure~\ref{fig:lf_lrg}) populates the highly luminous bright-end regime with a characteristically narrow distribution, while the fainter regime is simultaneously filled by the incoming ELG sample (Figure~\ref{fig:lf_elg}), which exhibits a markedly steeper faint-end trend. Meanwhile, the QSO sample (Figure~\ref{fig:lf_qso}) spans extremely bright absolute magnitudes ($M \sim -24$ to $-28$) but remains a negligible contributor to the overall space density baseline.

At $z > 1.0$, the combined sample is severely depleted by the $(1+z)^{-4}$ Tolman surface-brightness dimming and strict flux limits, leaving only high-luminosity QSOs and sparse, compact ELG outliers. Consequently, the global LF (Figure~\ref{fig:lf_combined}) becomes highly unstable and poorly constrained at the faint end. This empirical sequence illustrates how forcing a single parametric Schechter function onto the combined sample across $0.4 < z < 1.0$ yields an unphysical, artificial compromise between discrete tracer classes that populate independent regimes of the redshift-luminosity plane. These raw distributions therefore do not trace genuine galaxy evolution, but instead serve as a striking visual diagnostic of target-class mixing and coupled selection boundaries.

\subsection{Evolution of Schechter Parameters}

\begin{figure}[ht]
    \centering
    \includegraphics[width=0.4\textwidth]{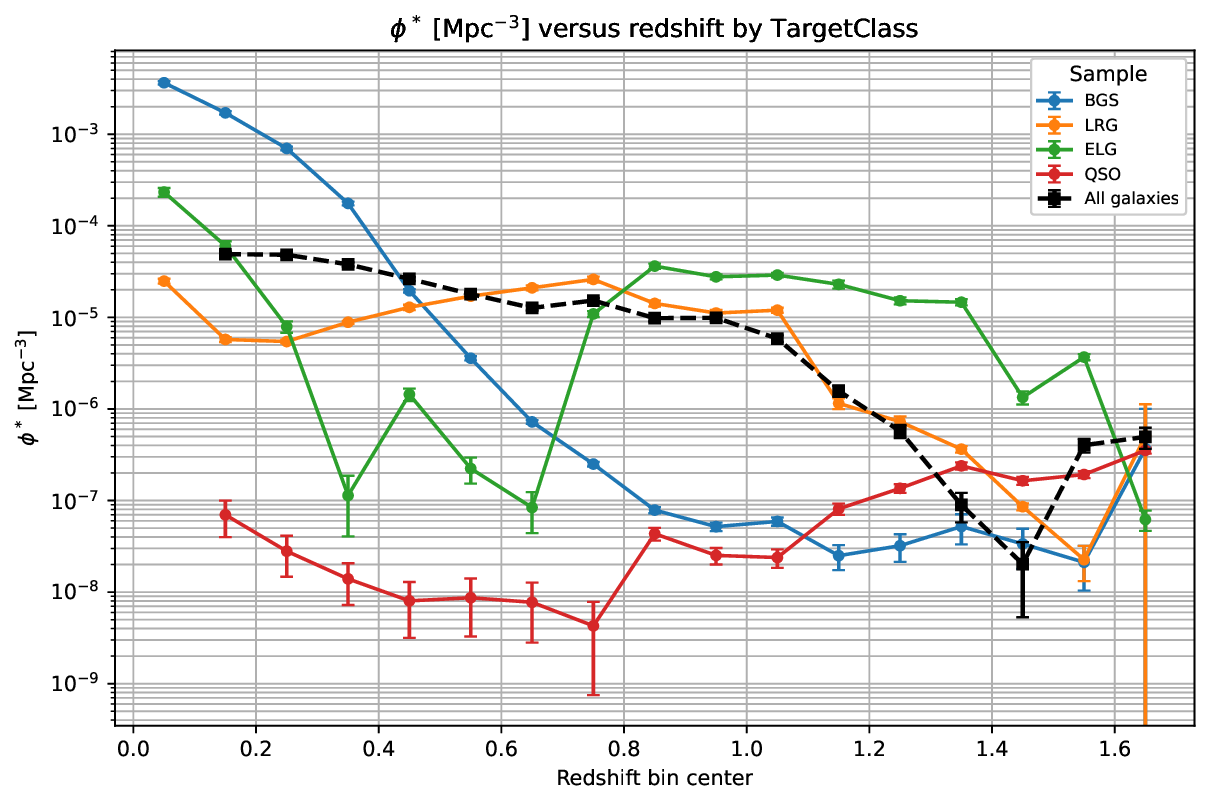}
   
    \caption{Diagnostic redshift evolution of the fitted Schechter density parameter, $\phi^*$ for the combined (global) DESI DR1 sample (black curve) and for the BGS (blue), LRG (orange), ELG (green), and QSO (red) target classes independently. The global $\phi^{*}$ declines by approximately four orders of magnitude between $z=0.2$ and $z=1.6$, with two prominent step-like drops near $z\approx0.6$ and $z\approx1.1$. These drops align with the effective volume limits of the BGS and LRG selection functions, respectively. The observed trajectories are presented strictly as an empirical diagnostic of survey incompleteness rather than a claimed physical evolution of the cosmic galaxy density. The reported uncertainties represent internal statistical errors propagated from $1/V_{\mathrm{max}}$ Poisson counting statistics and should be treated as lower bounds, as they omit cosmic variance and systematic selection uncertainties.}
    \label{fig:phi_star}
    

\end{figure}

\begin{figure}[ht]
    \centering
    \includegraphics[width=0.4\textwidth]{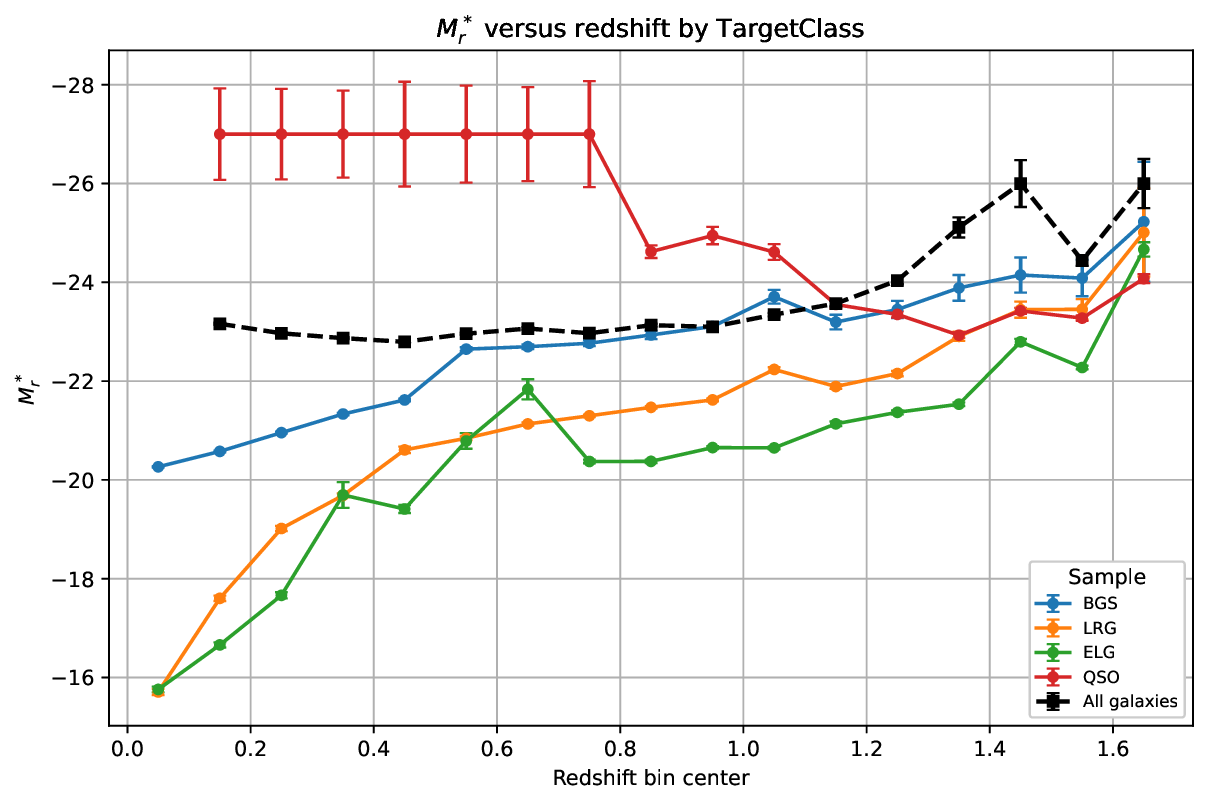}

    \caption{Diagnostic redshift trajectory of the fitted Schechter characteristic absolute magnitude, $M^*$ for the combined (global) DESI DR1 sample (black curve) and for the BGS (blue), LRG (orange), ELG (green), and QSO (red) target classes independently. The global $M^{*}$ remains approximately constant, $M^{*} \approx -23$, across $0.1<z<1.0$, reflecting the long-term dominance of massive galaxies in the combined sample. The BGS population exhibits a brightening from $\approx -20$ at $z\approx0.1$ to $\approx -23$ at $z>0.4$, driven by increasing LRG contamination. The LRG sample traces a fainter sequence ($M^{*} \approx -20$ to $-22$) across $0.4<z<1.0$, while ELGs trace the faintest sequence ($M^{*} \approx -20$ to $-23$) across $0.6 \lesssim z \lesssim 1.6$. These trends represent mathematical fits to target-selected, incomplete samples and must not be interpreted as intrinsic cosmic evolution of galaxy luminosities. Uncertainties shown account solely for Poisson counting statistics and represent lower limits on the true parameter uncertainties.}
    \label{fig:m_star}
    

\end{figure}

\begin{figure}[ht]
    \centering
    \includegraphics[width=0.4\textwidth]{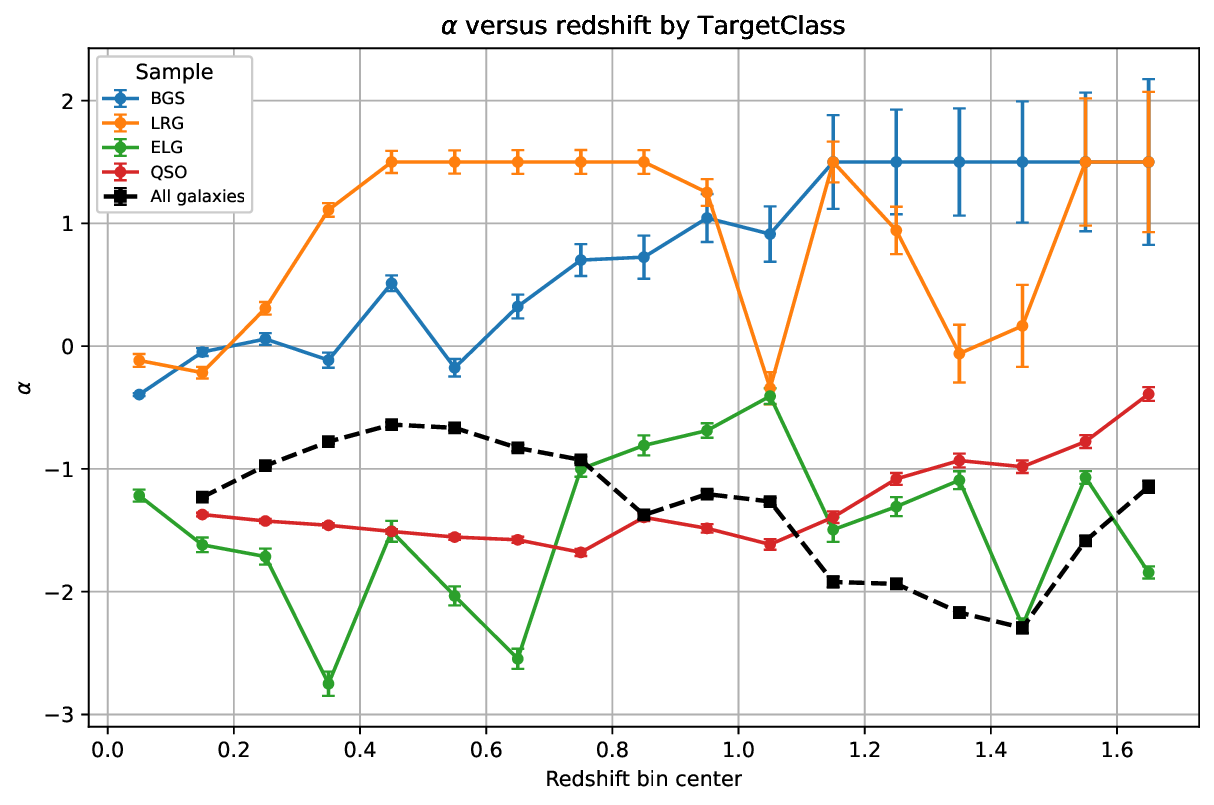}

    \caption{Diagnostic redshift behavior of the fitted Schechter faint-end slope parameter, $\alpha$ for the combined (global) DESI DR1 sample (black curve) and for the BGS (blue), LRG (orange), ELG (green), and QSO (red) target classes independently. The global $\alpha$ exhibits non-monotonic evolution: steep ($\alpha \approx -1$) at $z<0.4$, flattening to $\alpha \approx -0.8$ at $z\approx0.5$--$0.6$ (as the shallow-slope LRG population becomes dominant), and steepening again beyond $z\approx0.8$ (reaching $\alpha \approx -2.25$ at $z\approx1.5$, driven by compact star-forming galaxies and QSOs convolved with photometric uncertainties). At $z>1.2$, the global $\alpha$ becomes highly unstable as the sample is reduced to sparse outliers. The observed trajectories are presented strictly as an empirical diagnostic of survey incompleteness rather than a claimed physical evolution of the faint-end slope. The error bars reflect $1/V_{\mathrm{max}}$ Poisson statistical uncertainties and represent lower limits that do not capture systematic selection or cosmic variance uncertainties.}
    \label{fig:alpha}
    

\end{figure}

As established above, the evolution of the Schechter parameters ($\phi^*, M^*$, and $\alpha$) presented in Figures~\ref{fig:phi_star}--\ref{fig:alpha} must be interpreted strictly as fitted diagnostic quantities rather than intrinsic physical trends of the underlying galaxy population. Because the underlying samples are simultaneously shaped by redshift-dependent target-selection boundaries, cosmological attenuation effects (e.g., $(1+z)^{-4}$ Tolman dimming and K-corrections), observational flux/surface-brightness completeness limits (Malmquist bias), and instrumental constraints (such as fiber assignment collisions and aperture losses), these parameter trajectories serve as a direct visual and mathematical map of total survey incompleteness.

Figures~\ref{fig:phi_star},~\ref{fig:m_star}, and~\ref{fig:alpha} illustrate the redshift evolution of the three Schechter parameters — normalization density $\phi^{*}$, characteristic magnitude $M^{*}$, and faint-end slope $\alpha$ — for both the combined sample (black curve) and for each target class independently (BGS in blue, LRG in orange, ELG in green, QSO in red). We emphasize that the uncertainties associated with the best-fit Schechter parameters presented in Figures~\ref{fig:phi_star},~\ref{fig:m_star}, and~\ref{fig:alpha} are derived directly from the $1/V_{\mathrm{max}}$ Poisson statistical error budget. Because the underlying luminosity function measurement uncertainties omit systematic effects—such as cosmic variance, fiber assignment completeness fluctuations, and photometric zero-point uncertainties—the propagated uncertainties on these Schechter parameters represent statistical lower bounds and are systematically underestimated. Consequently, caution should be exercised against over-interpreting minor redshift variations or apparent parameter correlations, as these fits serve a diagnostic role in tracing sample boundaries rather than quantifying intrinsic astrophysical parameter scatter.

\paragraph{Normalization density $\phi^{*}$ (Figure~\ref{fig:phi_star}).} 
The global $\phi^{*}$ (black curve) declines by approximately four orders of magnitude between $z = 0.2$ and $z = 1.6$. This decline is punctuated by two prominent, non-physical step-like drops: the first near $z \approx 0.6$ (a factor of $\sim 10$) and the second near $z \approx 1.1$ (another factor of $\sim 10$). Rather than necessarily tracing a genuine cosmic depletion of the galaxy space density, these discontinuous drops align precisely with the effective volume limits and sharp selection boundaries of the BGS and LRG target classes, respectively. When examined by tracer, the BGS sample (blue) exhibits a high normalization density ($\phi^{*} \sim 10^{-3}$ to $10^{-4}$ Mpc$^{-3}$ mag$^{-1}$) at $z < 0.4$ before truncating sharply. Conversely, the LRG sample (orange) drops from $\phi^{*} \sim 10^{-5}$ at $z = 0.4$ to $\phi^{*} \sim 10^{-8}$ at $z = 1.6$, a steepening driven by both the strict LRG flux limits and uncorrected fiber collisions in high-density environments. The ELG sample (green) maintains a higher baseline ($\phi^{*} \sim 10^{-4}$) across $0.8 < z < 1.4$, reflecting the abundance of star-forming systems, until it is suppressed by the severe $(1+z)^{-4}$ Tolman surface-brightness dimming at higher redshifts. We note that the absolute vertical scale of all curves appears systematically depressed; this is a direct, expected consequence of omitting the geometric sky-area fraction of the DESI DR1 footprint in our diagnostic $1/V_{\mathrm{max}}$ estimator. Consequently, the overall behavior of $\phi^{*}$ in Figure~\ref{fig:phi_star} provides strong empirical proof that global LF normalization trends are structurally dominated by coupled survey selection boundaries and instrumental constraints rather than intrinsic galactic evolution.

\paragraph{Characteristic magnitude $M^{*}$ (Figure~\ref{fig:m_star}).} 
The global $M^{*}$ (black curve) remains approximately constant at $M^{*} \approx -23$ across the range $0.1 < z < 1.0$. Rather than reflecting a genuine physical feature of the combined population, this high luminosity plateau indicates that the uncorrected global sample is systematically anchored to the brightest, massive systems (primarily LRGs). In contrast, the class-separated BGS population (blue) exhibits a severe apparent evolution, brightening from $M^{*} \approx -20$ at $z \approx 0.1$ to $\approx -23$ for $z > 0.4$. This dramatic shift provides a clear empirical manifestation of the Malmquist bias and the $(1+z)^{-4}$ Tolman surface-brightness attenuation; as fainter systems systematically drop below detection limits at higher redshifts, the Schechter mathematical fit is artificially forced toward brighter characteristic magnitudes. Similarly, the stable but bright sequence tracked by the LRG ($M^{*} \approx -21$) and ELG samples at high redshifts is strongly influenced by the uncorrected Eddington bias, where photometric uncertainties scatter lower-luminosity systems into brighter magnitude bins. Therefore, the observed trajectories of $M^{*}$ in Figure~\ref{fig:m_star} do not depict intrinsic cosmic evolution, but instead serve as a diagnostic baseline illustrating how coupled observational selection effects deform the fundamental parameters of the galaxy luminosity function.

\paragraph{Faint-end slope $\alpha$ (Figure~\ref{fig:alpha}).} 
The global $\alpha$ (black curve) exhibits a severe, non-monotonic evolution across the analyzed redshift range. At $z < 0.4$, the global slope is entirely dictated by the BGS sample (blue), hovering around $\alpha \approx -1.0$, which is consistent with a normal faint end dominated by local star-forming galaxies. Between $z = 0.4$ and $z = 0.8$, the global $\alpha$ flattens significantly, reaching values of $\alpha \approx -0.8$ at $z \approx 0.5$--$0.6$. This artificial flattening is driven by the sudden dominance of the LRG population (orange), which exhibits an anomalous, inverted, or highly positive slope up to $\alpha \approx +1.5$ across $0.4 < z < 1.0$. Rather than a genuine physical feature, this positive $\alpha$ is a direct artifact of the strict low-luminosity cuts in the LRG target selection combined with uncorrected fiber collisions, which systematically deplete lower-luminosity galaxies in high-density environments. The ELG sample (green) maintains a steeper baseline ($\alpha \approx -1.0$ to $-1.5$) over $0.6 < z < 1.6$, but becomes highly unstable and unconstrained due to the progressive loss of diffuse, extended systems to the severe $(1+z)^{-4}$ Tolman surface-brightness dimming. Remarkably, beyond $z \approx 0.8$, the global $\alpha$ appears to steepen again, reaching non-physical values near $\alpha \approx -1.25$ at $z \approx 1.0$ and plummeting toward $\alpha \approx -2.25$ at $z \approx 1.5$. This high-redshift steepening is a mathematical artifact of the Schechter fit compensating for severe completeness losses; as the bulk of the typical galaxy population drops below the detection threshold, the convolution of increasing photometric uncertainties with the remaining sparse, compact star-forming systems and high-luminosity QSOs (red) mimics a highly steepened faint-end distribution. Consequently, the erratic, non-monotonic behavior of the global $\alpha$ in Figure~\ref{fig:alpha} serves as a clear diagnostic signature of how coupled instrumental selection effects, Tolman dimming, and Eddington-like photometric scattering masquerade as rapid cosmic evolution in uncorrected survey data.

\subsection{Diagnostic Baseline and Necessary Corrections}

The analysis presented in Sections~\ref{sec:lf_analysis} uses standard DESI pipeline corrections for Galactic extinction, $K$-corrections, and photometric normalization. 

A robust correction for structural selection bias requires: (i) rest-frame surface-brightness estimates $\mu_{\mathrm{corr}}$ following Equation~(\ref{eq:mucorr}), and (ii) a redshift-independent rest-frame surface-brightness limit $\mu_{\mathrm{lim,rest}}$ applied consistently across all bins. This approach, outlined in Section~\ref{subsec:aperture_extinction_restframe}, is necessary to stabilize $\alpha$ against Tolman-driven dropouts and isolate the genuine evolutionary sequence from the superimposed effects of target-class transitions. The diagnostic baseline presented in Figures~\ref{fig:lf_combined}--\ref{fig:alpha} explicitly illustrates the magnitude of the distortion, verifying that omitting the rigorous multi-dimensional boundary layer leads directly to the systematic instabilities described above.



\subsection{Conceptual Framework for Downstream Analysis Mitigations}
\label{sec:mitigation_framework}

The conceptual framework for downstream analysis mitigations outlined herein is designed to address the dominant observational selection effects affecting the DESI DR1 galaxy sample. Rather than applying ad-hoc empirical corrections directly to the data, this diagnostic methodology establishes a clear pathway to account for selection boundaries, completeness limits, and surface-brightness constraints in future cosmological pipelines. Nevertheless, several methodological limitations must be explicitly considered when incorporating these diagnostics into downstream analyses.

First, the proposed framework does not constitute a full multidimensional inversion of the complete DESI selection function, nor does it explicitly incorporate forward-modeling techniques based on realistic mock catalogs. Residual systematics associated with target-class overlap, sub-dominant photometric uncertainties, environmental dependencies, and cosmic variance may therefore persist.

Second, the corrections and constraints described here are intended to drastically reduce the impact of known observational trends rather than completely eliminate all sources of error. Consequently, the resulting sample configurations should be regarded as substantially less biased approximations of the underlying galaxy population, rather than perfectly unbiased realizations of galaxy evolution.

Third, it is worth noting that the Schechter parameter evaluations utilized as a diagnostic baseline in Section~\ref{sec:lf_analysis} did not yet enforce the strict empirical surface-brightness threshold cuts from Equation~(\ref{eq:mulim}). This diagnostic baseline acts as the control sample that explicitly illustrates the magnitude of the distortion, verifying that omitting the rigorous multi-dimensional boundary layer leads directly to the systematic instabilities shown in Figures~\ref{fig:phi_star},~\ref{fig:m_star}, and~\ref{fig:alpha}.

Future work combining DESI observations with survey simulations and detailed selection-function modeling will be necessary to quantify the remaining systematic uncertainties and further refine evolutionary measurements. In particular, forward-modeling approaches that inject realistic galaxy populations into mock DESI catalogs offer a promising path toward a full inversion of the survey selection function.

\section{Discussion}
\label{sec:discussion}

The principal contribution of this work is not simply confirming the inherent incompleteness of the DESI spectroscopic sample—an operational feature already established in survey design literature \citep{Myers2023}—but rather providing the first quantitative identification, visual mapping, and analytical framework for the ``fork'' geometry emerging in the $z$--$M$ plane. By systematically dissecting this distribution across target classes, we isolate the precise root causes driving these structural features: the physical intersection of DESI target-selection color boundaries, $1/V_{\mathrm{max}}$ volume limits, quasar/AGN central emission contributions at high redshifts, and cosmological $(1+z)^{-4}$ Tolman surface-brightness dimming acting against sky-background thresholds.

Crucially, our $1/V_{\mathrm{max}}$ Schechter parametric fits act as a strictly diagnostic tool rather than a direct empirical correction. By retaining incomplete boundary bins and evaluating the raw observational data against parametric models, this work establishes a quantitative baseline that maps operational selection limits and demonstrates where completeness breaks down. As detailed in Section~\ref{sec:lf_analysis}, the severe, high-significance modulations observed relative to smooth Schechter fits reflect the sharp operational boundaries of target-selection algorithms and fiber-assignment limits rather than intrinsic galaxy population evolution.

The results presented here should not be interpreted as evidence that the observed DESI fork geometry is entirely artificial. Rather, the observed structure emerges from the interaction between intrinsic galaxy bimodality \citep{Strateva2001, Baldry2004} and the observational selection effects imposed by the survey design. The physical distinction between star-forming and quiescent systems remains real, while the DESI targeting strategy modifies the manner in which these populations are sampled and represented in the $z$--$M$ plane.

The existence of two dominant galaxy populations—the star-forming Blue Cloud and the quiescent Red Sequence—is a well-established physical characteristic of the galaxy population. In an ideal volume-complete survey, these populations would occupy a continuous but bimodal distribution in luminosity, color, and stellar mass, connected through the intermediate Green Valley \citep{Faber2007, Schawinski2014}. However, the DESI targeting strategy intentionally prioritizes different classes of objects over distinct redshift intervals. As a consequence, the observed galaxy distribution inherits discontinuities that are not physically associated with galaxy transformation processes.

Our analysis shows that the apparent depletion of galaxies within the region separating the two branches of the fork cannot be interpreted as direct evidence for accelerated quenching or a rapidly evolving Green Valley population. Instead, the combined action of magnitude limits, color-selection boundaries, fiber-assignment priorities, photometric uncertainties, and surface-brightness incompleteness \citep{Blanton2003, Driver2005, Johnston2021} artificially amplifies the contrast between the two major galaxy populations. In this sense, the observed gap is not equivalent to the physical Green Valley, although both partially overlap within the same parameter space.

The situation becomes particularly important near $z \sim 0.7$, where the two branches appear to converge. A superficial interpretation could suggest a physical disappearance of galaxy bimodality at intermediate redshifts. However, our results indicate that this convergence is largely driven by the increasing overlap between the BGS, LRG, ELG, and QSO selection functions, as demonstrated by the target-class mapping in Figure~\ref{fig:TargetClasseSMr} and the surface-brightness analysis in Figures~\ref{fig:z_vs_M_panels}. The observed merging therefore reflects the transition between different target populations rather than a sudden modification of the underlying galaxy population.

The analysis of surface-brightness distributions further reinforces this interpretation. The strong dependence of galaxy detectability on effective surface brightness, combined with cosmological Tolman dimming, preferentially removes diffuse systems from the observed sample at increasing redshift. This effect is particularly relevant for extended star-forming galaxies and low-surface-brightness systems, which progressively migrate below the survey detection threshold. Consequently, part of the apparent evolution of the galaxy distribution in the $z$--$M$ plane is produced by structural incompleteness rather than genuine physical evolution.

An important implication is that DESI DR1 should not be treated as a statistically homogeneous galaxy sample. While the survey provides an unprecedented spectroscopic census of the low- and intermediate-redshift Universe, its primary optimization for cosmological tracers inevitably introduces complex selection boundaries that must be explicitly modeled. Failure to account for these boundaries can lead to biased estimates of luminosity evolution, stellar-mass growth, quenching rates, structural evolution, and the relative abundance of galaxy populations.

The conceptual framework for downstream mitigations proposed in this work provides a practical pathway for accounting for these effects. By combining luminosity completeness limits, surface-brightness constraints, aperture normalization, extinction corrections, and careful redshift binning, it becomes possible to isolate regions of parameter space where evolutionary trends can be interpreted with substantially greater confidence. Although no mitigation scheme can fully eliminate all survey-dependent systematics, the methodology presented here significantly reduces the risk of confusing instrumental footprints with genuine astrophysical signals.

Finally, the conclusions reached here extend beyond DESI itself. Many modern cosmological surveys employ tracer-dependent target selection strategies optimized for specific scientific goals. As future spectroscopic datasets continue to increase in size and complexity, understanding how survey design reshapes the observed galaxy population will become increasingly important. The DESI fork geometry therefore provides a useful case study illustrating how instrumental selection effects can mimic or distort evolutionary signatures if not properly characterized.

Although the diagnostic framework proposed here has the potential to substantially reduce the dominant observational biases affecting DESI DR1, residual systematics associated with target selection, photometric uncertainties, and structural measurements inevitably remain. Future studies based on realistic mock catalogs and forward-modeling approaches will be necessary to quantify the full impact of the DESI selection function on galaxy-evolution analyses.

\section{Conclusions}\label{sec:conclusions}

Using the DESI DR1 spectroscopic sample and its associated Legacy Survey photometry, we have investigated the origin and implications of the characteristic ``fork'' geometry observed in the $z$-$M$ plane. Our main conclusions are summarized as follows:

\begin{enumerate}

    \item The fork-like distribution observed in DESI DR1 cannot be interpreted solely as a physical manifestation of galaxy evolution. Instead, it arises from the superposition of intrinsic galaxy bimodality \citep{Baldry2004} and the sharp, redshift-dependent targeting boundaries of DESI \citep{Myers2023, Hahn2023, Zhou2023, Raichoor2023, Chaussidon2023}, a systematic feature common to target-selected spectroscopic surveys \citep{Blanton2003, Driver2005}.

    
    \item The apparent observational vacuum separating the two main branches at $z \lesssim 0.7$ should not be interpreted as the physical Green Valley. Instead, it is significantly amplified by target-selection boundaries, flux incompleteness, surface-brightness limitations, and photometric selection effects.
    
    \item The convergence of the two branches near $z \sim 0.7$ is largely driven by the increasing overlap of the BGS, LRG, ELG, and QSO target classes. Consequently, the observed merging of populations at intermediate redshift cannot be attributed solely to intrinsic galaxy evolution.

    \item Surface-brightness selection effects and cosmological Tolman dimming play a major role in shaping the observed galaxy distribution in $z$--$M$--$\mu$ space. These effects preferentially remove diffuse systems from the sample and contribute to the apparent evolution of the galaxy population across redshift.
    
    \item The combined influence of Malmquist bias, photometric uncertainties, aperture effects, and structural incompleteness can substantially distort measurements of galaxy evolution if not explicitly controlled.

    \item We summarize a set of observational considerations involving extinction corrections, $K$-corrections, aperture normalization, inclination effects, completeness limits, and surface-brightness constraints. Together, these elements provide a conceptual mitigation framework for constructing statistically robust subsamples suitable for downstream evolutionary studies.

\end{enumerate}

In addition to these scientific conclusions, we provide the following practical recommendation for researchers utilizing DESI DR1:

\begin{quote}
    For studies focused on galaxy formation and evolution, we suggest defining scientific samples within volume-limited and surface-brightness-complete regions of parameter space, using DR10 Legacy Surveys photometry for sample definition and luminosity-based analyses, while employing DR1 spectroscopy primarily for precise redshift measurements and spectroscopic diagnostics.
\end{quote}

Finally, we emphasize that the DESI DR1 dataset remains an exceptionally powerful resource for extragalactic astronomy. However, its scientific exploitation requires careful treatment of the survey selection function to ensure that observed trends reflect genuine astrophysical processes rather than artifacts introduced by the observational strategy.

In summary, the DESI fork geometry represents a vivid example of how survey design can imprint artificial structures onto the observed galaxy population. Properly accounting for these effects is essential for transforming the unprecedented statistical power of DESI into reliable constraints on galaxy evolution across cosmic time.

The diagnostic framework proposed here should therefore be viewed as a practical tool for mapping and minimizing the dominant observational biases present in DESI DR1, rather than as a complete inversion of the survey selection process. Its primary value lies in enabling more robust statistical studies of galaxy evolution while highlighting the importance of carefully accounting for survey design when interpreting large spectroscopic datasets.

\begin{acknowledgments}
We thank the DESI collaboration for the DR1 public release. We also thank the Instituto de Astronomía y Meteorología (UdG, México) for all the facilities provided for the realization of this project. A.N.-N. acknowledges support from CONAHCyT and PRODEP (México). P.L. gratefully acknowledges support by the GEMINI ANID project No. 32240002. The work of R.J.D. is supported by the International Gemini Observatory, a program of NSF NOIRLab, which is managed by the Association of Universities for Research in Astronomy (AURA) under a cooperative agreement with the U.S. National Science Foundation, on behalf of the Gemini partnership of Argentina, Brazil, Canada, Chile, the Republic of Korea, and the United States of America.
\end{acknowledgments}



\end{document}